\documentclass[twocolumn, apl, amsmath, amssymb, superscriptaddress]{revtex4-2}
\usepackage{epsf}      
\usepackage{graphicx}
\usepackage{color}
\usepackage{soul}
\usepackage{gensymb}
\usepackage{sidecap}
\usepackage{amsmath}
\usepackage{mathtools}
\usepackage[table]{xcolor}
\usepackage[colorlinks=true, linkcolor=blue, citecolor=blue, urlcolor=blue]{hyperref}

\begin{document}

\title{Magnetic field induced modification of a first-order ferromagnetic transition in Eu$_2$In}
\author{Ajay Kumar}
\email{ajay1@ameslab.gov}
\affiliation{Ames National Laboratory, U.S. Department of Energy, Iowa State University, Ames, Iowa 50011, USA} 
 \author{Anis Biswas}
\affiliation{Ames National Laboratory, U.S. Department of Energy, Iowa State University, Ames, Iowa 50011, USA} 
 \author{Trevor A. Tyson}
\affiliation{Department of Physics, New Jersey Institute of Technology, Newark, NJ 07102, USA}
 \author{Daniel Haskel}
\affiliation{Advanced Photon Source, Argonne National Laboratory, Lemont, Illinois 60439, USA} 
 \author{Christopher J. Pollock}
\affiliation{Cornell High Energy Synchrotron Source, Wilson Laboratory, Cornell University, Ithaca, NY 14853, USA}
 \author{Yaroslav Mudryk}
\affiliation{Ames National Laboratory, U.S. Department of Energy, Iowa State University, Ames, Iowa 50011, USA} 

\date{\today}

\begin{abstract}

We present a comprehensive study of the temperature- and magnetic-field-dependent magnetization, specific heat, and local crystal structure across the first-order ferromagnetic--paramagnetic transition in Eu$_2$In. Anomalies in the magnetocaloric response are observed near $H \approx 25$~kOe, including changes in field scaling of magnetic entropy, local entropy exponent, and universal master curve, which suggest an apparent weakening of the first-order character of the transition. However, quantitative analysis of the magnetocaloric parameters together with modified Arrott plots demonstrates that the transition remains first order up to at least 70~kOe. Specific-heat measurements reveal a field-induced splitting of the sharp zero-field anomaly into a doublet, providing a natural explanation for the change in the magnetocaloric response. Magnetic field dependent extended x-ray absorption fine structure (EXAFS) measurements show no detectable field-induced changes in the local coordination environment of Eu. We therefore attribute these observations to a magnetic field induced two-step transition process in Eu$_2$In.

\end{abstract}

\maketitle

\section{\noindent ~Introduction}

A fundamental understanding of the thermodynamic nature of phase transitions is essential for elucidating the non-trivial responses of materials to applied magnetic fields, which underpin a wide range of magnetofunctional properties \cite{Hou_NRM_22, Kainuma_Nat_06}. In particular, first-order magnetic phase transitions (FOMPTs), characterized by their discontinuous nature, often exhibit strong non-linear responses to external stimuli and are therefore of considerable interest for functional applications \cite{Rivadulla_PRL_03, Roch_PRL_20, Pecharsky_PRL_97}. Giant magnetocaloric effects (GMCE), for example, are typically observed in systems undergoing discontinuous magnetic or magnetostructural transitions \cite{Pecharsky_PRL_97, Oliveira_PRB_02, Pecharsky_PRL_03, Kumar_JAP_24}. A prototypical case is Gd$_5$(Si$_2$Ge$_2$), which exhibits a pronounced GMCE near room temperature associated with a first-order phase transition (FOPT) \cite{Pecharsky_PRL_97, Choe_PRL_2000, Mozharivskyj_JACS_05}. In many materials, such transitions originate from strong coupling between magnetic order and lattice degrees of freedom,  that are inherently accompanied by thermal hysteresis. Although extensive efforts over the past decades have focused on understanding the thermodynamic processes occurring in the vicinity of FOPTs \cite{Piazzi_PRA_17, Erbesdobler_JAP_20, Bennati_APL_20, Guillou_SM_19, Hardy_AM_22}, hysteresis associated with the transitions remains a major challenge, as it leads to energy dissipation, reduces refrigeration efficiency, and limits long-term cycling stability in solid-state caloric applications \cite{Shamberger_PRB_09, Provenzano_Nature_04}.

In this context, R$_2$In (R = rare earth) compounds have recently attracted particular attention \cite{Biswas_JPD_26}, as certain members of this family exhibit large magnetic entropy changes ($\Delta S_{\rm M}$) under modest magnetic fields, negligible thermal hysteresis, and minimal structural distortion at the first-order ferromagnetic (FM) to paramagnetic (PM) transition—characteristics highly desirable for practical magnetocaloric applications \cite{Guillou_NC_18, Biswas_PRM_22, Biswas_PRB_20, Liu_APL_21, Cui_JMCT_22, Forker_PRB_05, GegenMater25}. Within the R$_2$In family, compounds with R = Eu and Pr exhibit first-order transitions \cite{Guillou_NC_18, Biswas_PRB_20}, R = Nd displays a borderline behavior between first- and second-order transitions \cite{Biswas_PRM_22, Liu_APL_21, GegenMater25}, while the remaining members undergo either second-order or no  magnetic transitions \cite{Forker_PRB_05, Seale_JAP_79, Biswas_JPD_26}. R$_2$In compounds crystallize in a Co$_2$Si-type orthorhombic structure for R = Eu and Yb \cite{Guillou_NC_18, Guillou_PRM_20}, whereas all other members adopt the Ni$_2$In-type hexagonal structure \cite{Baela_JLCM_88, Biswas_JPD_26}, indicating no direct connection between the crystal structure and the order of the magnetic transition in these compounds. However, the R = Eu and Yb members host divalent rare-earth ions \cite{Ritter_JalCom_24, Ryan_AIP_19, Guillou_PRM_20}, whereas in all other R$_2$In compounds the rare-earth elements occur in the trivalent state, suggesting an intricate correlation between the valence state and the crystal structure of these compounds.

In this family, Eu$_2$In exhibits the highest magnetic entropy change of $\approx$~$-28$ J/kg-K for $\Delta H = 20$ kOe, with negligible thermal hysteresis (0.1~K) and a minimal change of $\approx$ 0.1\% in the unit cell volume across its first-order FM--PM transition at $T_{\rm c} = 55$ K \cite{Guillou_NC_18}. Theoretical calculations demonstrate that the strong hybridization between Eu-5$d$ and In-5$p$ states near the Fermi level gives rise to long-range exchange interactions between Eu atoms, thereby  triggering the ferromagnetic transition, which in turn underlies the large magnetic entropy change at low magnetic fields in this compound \cite{Guillou_NC_18}. \textit{Ab initio} calculations by Tapia \textit{et al.} attribute the discontinuous phase transition and the associated large latent heat in Eu$_2$In to a purely electronic mechanism arising from the divalent nature of the Eu ions in the compound \cite{Tapia_PRB_20}. Subsequently, Alho \textit{et al.} employed a model Hamiltonian—independent of the details of the electronic structure—that incorporates magnetic exchange and magnetoelastic interaction terms, and successfully reproduces the magnetic-field dependence of $T_{\rm c}$ as well as the temperature dependence of the magnetization and magnetocaloric properties \cite{Alho_PRB_20}. Furthermore, by analyzing the evolution of mathematical critical points in the magnetic free energy as a function of magnetization with respect to temperature and applied field, the authors identified a critical magnetic field $H_{\rm c} \approx 25$~kOe, above which the transition in Eu$_2$In thermodynamically changes from first order to second order \cite{Alho_PRB_20}. This theoretically predicted change in the order of the magnetic transition is of particular relevance for magnetocaloric applications, since a reduction in the discontinuity and sharpness of the transition at higher fields can markedly influence the magnetocaloric performance. Nevertheless, no experimental investigations have yet been conducted to verify this proposed field-induced crossover in Eu$_2$In.

In this study, we report a detailed investigation of the temperature- and field-dependent magnetization, specific heat, and local crystal structure across the first-order FM–PM transition in Eu$_2$In. Distinct anomalies are identified in the magnetocaloric response, including the maximum magnetic entropy change ($\Delta S_{\rm M}^{\rm max}$), the local power exponent ($n$), and the universal scaling behavior, around $H\sim25$ kOe—the crossover field predicted in Ref. \cite{Alho_PRB_20}. However, a quantitative analysis of the extracted parameters, together with the modified Arrott plot analysis, demonstrates that the transition retains its first-order nature up to the highest applied field of 70 kOe. The specific heat measurements reveal that the sharp zero-field anomaly evolves into a well-defined doublet with increasing magnetic field, while extended x-ray absorption fine structure (EXAFS) measurements show no detectable field-induced modifications in the local coordination environment within the experimental resolution. This behavior is discussed in terms of the possible emergence of a field-driven two-step process associated with the transition.

\section{\noindent ~Experimental}

 Polycrystalline Eu$_2$In was synthesized from stoichiometric amounts of high-purity indium (99.9995\%, Alfa Aesar) and europium metal (99.995\%, Materials Preparation Center, Ames National Laboratory). The elements (total mass $\sim$3~g) were loaded into a tantalum (Ta) crucible and sealed by arc welding under an ultrapure argon atmosphere. Prior to sample loading, the Ta crucible was degassed under high vacuum ($\sim 10^{-6}$ Torr) at 1800~$^\circ$C for 2~h. The Ta crucible was then sealed in a quartz tube under high-purity He gas to prevent oxidation of the crucible at elevated temperatures. The sample was heated to 900~$^\circ$C for 12~h in a resistive furnace, flipped, and reheated three times to improve homogeneity, and then annealed at 650~$^\circ$C for 24~h. After furnace cooling, the sample was extracted from the crucible and handled in an Ar-filled glovebox due to its high reactivity with air.

Room-temperature powder x-ray diffraction (XRD) data were collected using a Rigaku diffractometer with Mo $K_{\alpha}$ radiation \cite{Holm_RSI_04}. For the XRD measurements, the sample was ground using a mortar and pestle and sieved through a 36~$\mu$m mesh inside an Ar-filled glovebox. The powdered sample was mounted in a shallow circular recess on a flat glass holder and sealed between two Kapton tapes inside the glovebox. The Rietveld refinement of the XRD data was performed using the FullProf Suite, with a pseudo-Voigt peak shape and linear interpolation between background points.

Temperature- and field-dependent magnetization measurements were performed using a superconducting quantum interference device (SQUID) magnetometer (MPMS XL-7, Quantum Design, USA). The sample was loaded into an airtight holder inside the glovebox to prevent air exposure. Magnetization isotherms at successive temperatures were recorded by increasing the magnetic field from 0 to 70~kOe in a stable (no-overshoot) mode, followed by decreasing the field from 70~kOe to 500~Oe linearly and from 500~Oe to 0~Oe in an oscillatory mode before proceeding to the next temperature, taking into account the low coercivity of the sample ($\sim$50~Oe). Specific-heat measurements were carried out using a Physical Property Measurement System (PPMS, Quantum Design, Inc.) employing both the standard 2$\tau$ relaxation technique and the long heat-pulse method \cite{Hardy_JPCM_09, Lashley_Cryo_43}.

Temperature- and field-dependent x-ray absorption spectroscopy (XAS) measurements at the Eu $L_{3}$ edge were performed at the PIPOXS beamline (ID2A) of the Cornell High Energy Synchrotron Source (CHESS) using the NJIT 10~T superconducting Oxford Spectromag4000-10 magnet system. Eu$_2$In powder was ground and homogenized with boron nitride (BN) in a fixed ratio and pressed into a 13~mm-diameter pellet inside an inert-atmosphere glovebox. The pellet was sealed between two Kapton tapes to prevent air exposure. The sample was cooled in He vapor. At each temperature and magnetic field, four spectra were collected and averaged. Energy calibration was performed using a simultaneously measured Eu metal foil reference to correct for possible energy shifts between scans \cite{Bunker_Book_10, Kumar_PRB_22}. X-ray absorption near-edge structure (XANES) and  EXAFS data reduction and analysis were performed using Athena and Artemis software packages (Demeter suite) \cite{Ravel_JSR_05}. Background subtraction and normalization were carried out in Athena, and EXAFS fitting was performed in  Artemis using theoretical scattering paths calculated with \textsc{FEFF6}.

\section{\noindent ~Results}

\subsection{\noindent ~X-ray diffraction}

 \begin{figure*}
\includegraphics[width=1\textwidth]{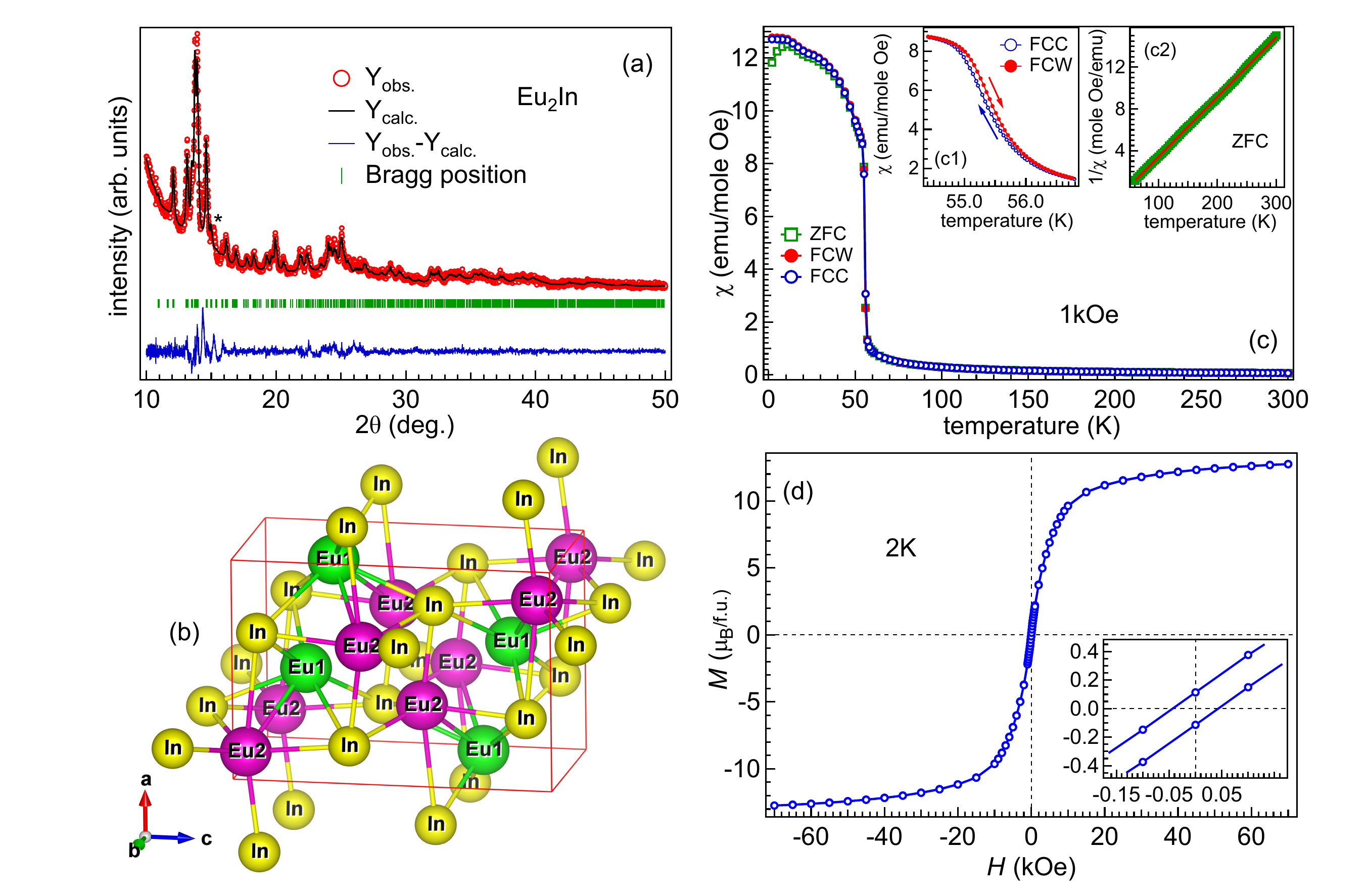} 
\caption{(a) Rietveld refinement of the room-temperature powder XRD pattern of Eu$_2$In collected using Mo $K_{\alpha}$ radiation. Red symbols represent the observed intensities, and the black solid line denotes the calculated profile. Vertical tick marks indicate the Bragg positions corresponding to the orthorhombic $Pnma$ (Co$_2$Si-type) phase. The bottom trace shows the difference between the observed and calculated patterns. An asterisk indicates the presence of a small amount ($\sim$2\%) of a secondary Eu$_8$In$_3$ phase. (b) Three-dimensional representation of the crystal structure of Eu$_2$In, highlighting the two inequivalent Eu sites. (c) Temperature-dependent magnetic susceptibility ($\chi$--$T$) measured in zero field cooled (ZFC), field cooled warming (FCW), and field cooled cooling (FCC) protocols at   1~kOe in the temperature stable mode. Inset (c1) shows high resolution FCC and FCW data recorded in the temperature sweep mode in the vicinity of $T_{\rm c}$. Inset (c2) shows the Curie--Weiss fit to the FCW data between 60 to 300~K. (d) Field-dependent magnetization measured at 2~K. The inset shows an enlarged view of the low-field region.}
\label{Fig1}
\end{figure*}

\begin{table*}
    \centering
    \caption{Room-temperature structural parameters and refinement quality indicators of Eu$_2$In obtained from Rietveld refinement of powder XRD data collected using Mo $K_{\alpha}$ radiation. Values in parentheses represent the standard deviations of the last significant digits. }
  
    \renewcommand{\arraystretch}{1.25}
    \setlength{\tabcolsep}{7pt}

    \begin{tabular}{cc|cccccc|c}
        \hline \hline
        Lattice parameters  &  & Atomic position & site & x &  y & z &  & Fitting parameters  \\
        \hline
        $a$ (\AA) & 7.4507(7) &  Eu1 & 4$c$ & 0.0322(10)  & 0.25  & 0.7074(6) &  & $\chi^2 =$ 1.96 \\
        $b$ (\AA) & 5.5795(5) &   Eu2 & 4$c$  & 0.1802(7) & 0.25 &  0.0715(6) &  & $R_p$ = 5.11 \\
        $c$ (\AA) & 10.3128(10) &  In  &  4$c$ & 0.2380(12) &  0.25  & 0.3926(7) &  & $R_{wp}$ = 7.39 \\
        cell volume (\AA$^3$) & 428.72(7) &    &  &  &   &  &  &  \\
        \hline
        \hline
    \end{tabular}

\label{T_critical}
\end{table*}

The room-temperature powder XRD pattern of Eu$_2$In is shown in Fig.~\ref{Fig1}(a). Rietveld refinement confirms that Eu$_2$In crystallizes in the Co$_2$Si-type orthorhombic structure (space group $Pnma$), in agreement with previous reports \cite{Guillou_NC_18,Ritter_JalCom_24}. However, careful inspection reveals a small amount ($\sim$2\%) of a secondary Eu$_8$In$_3$ phase, indicated by an asterisk in Fig.~\ref{Fig1}(a) (not included in the refinement). This trace amount of the secondary phase in Eu$_2$In is not uncommon and has also been observed previously \cite{Guillou_PRM_20}. In the present study we focus primarily on field-induced changes in the nature of the transition in Eu$_2$In around $T_{\rm c} = 55$~K; therefore, this minor secondary phase is not expected to influence the conclusions of our findings. The increasing background at low $2\theta$ values ($\lesssim 20^\circ$) arises from the Kapton tape used to protect the sample from air exposure. The refined lattice parameters, atomic Wyckoff positions, and refinement quality indicators are summarized in Table~1. Figure~\ref{Fig1}(b) shows a three-dimensional representation of the Eu$_2$In crystal structure, highlighting the two crystallographically inequivalent Eu sites.

\subsection{\noindent ~Magnetization}

The temperature-dependent magnetic susceptibility, $\chi = M/H$, of Eu$_2$In measured at an applied field of $H = 1$~kOe under zero field cooled (ZFC), field cooled warming (FCW), and field cooled cooling (FCC) protocols is shown in Fig.~\ref{Fig1}(c). A sharp magnetic transition with no apparent thermal hysteresis is observed at $T_{\mathrm{c}} \approx 55$~K, consistent with the previously reported first-order ferromagnetic to paramagnetic transition \cite{Guillou_NC_18}. However, the ``temperature settle'' mode of the magnetization measurement is not suitable for resolving the small thermal hysteresis associated with a first-order transition. This limitation arises because even a slight overshoot of the target temperature can drive the system across the phase boundary, transforming it from one phase to the other; upon returning to the set temperature, the system may remain trapped in the transformed state. Therefore, to reliably capture the hysteretic behavior near $T_{\mathrm{c}}$, magnetization measurements were performed upon both heating and cooling using the ``temperature sweep'' mode with a sweep rate of 0.1~K/min, as shown in the inset (c1) of Fig.~\ref{Fig1}(c). A clear thermal hysteresis of $\sim$0.1~K is observed at $T_{\mathrm{c}}$, confirming the first-order nature of the transition. Furthermore, the ZFC susceptibility was fitted to the Curie--Weiss law, $\chi = C/(T - T_\theta)$, where $C$ is the Curie constant and $T_\theta$ is the Curie--Weiss temperature. The linear fit of $1/\chi$ versus $T$ over the temperature range 60--300~K, shown in the inset (c2) of Fig.~\ref{Fig1}(c), yields $T_\theta = 40.1$~K and an effective magnetic moment of $\mu_{\mathrm{eff}} = 8.36~\mu_B$/Eu. The fact that $T_\theta < T_{\mathrm{c}}$ is attributed to the very sharp first-order character of the transition. The extracted $\mu_{\mathrm{eff}}$ is slightly larger than the free-ion value expected for Eu$^{2+}$ ($7.94~\mu_B$), which may indicate a small additional contribution from exchange polarization of itinerant conduction band states (with Eu $5d$ character) by the localized Eu $4f$ moments, rather than a purely localized $4f$ response. Figure~\ref{Fig1}(d) shows the field dependence of the magnetization measured at 2~K. The magnetization does not fully saturate even up to $H = 70$~kOe. At 70~kOe, the moment reaches $\sim 12.7~\mu_B$/f.u., which is lower than the value expected for complete ferromagnetic alignment of two Eu$^{2+}$ ions ($g_J J = 7~\mu_B$/Eu). The hysteresis loop exhibits a small coercive field of $\sim 50$~Oe, as highlighted in the inset showing an enlarged view of the low-field region. Overall, Eu$_2$In combines minimal thermal and magnetic hysteresis with a pronounced change in magnetization at $T_{\mathrm{c}}$, characteristics that are favorable for magnetocaloric applications \cite{Guillou_NC_18,Pecharsky_PRL_97}.

To investigate the magnetic-field-induced changes in the nature of this transition in Eu$_2$In, the virgin magnetization isotherms have been recorded across $T_{\rm c}$ with a temperature step of $\Delta T$ = 0.5~K [see Fig. \ref{Fig10}(a) of the Appendix]. The magnetic entropy change ($\Delta S_{\rm M}$) is then calculated from these isotherms by employing the Maxwell thermodynamic relation  \cite{Phan_JMMM_07, Kumar_PRB_20}

\begin{figure}
\includegraphics[width=3.5in]{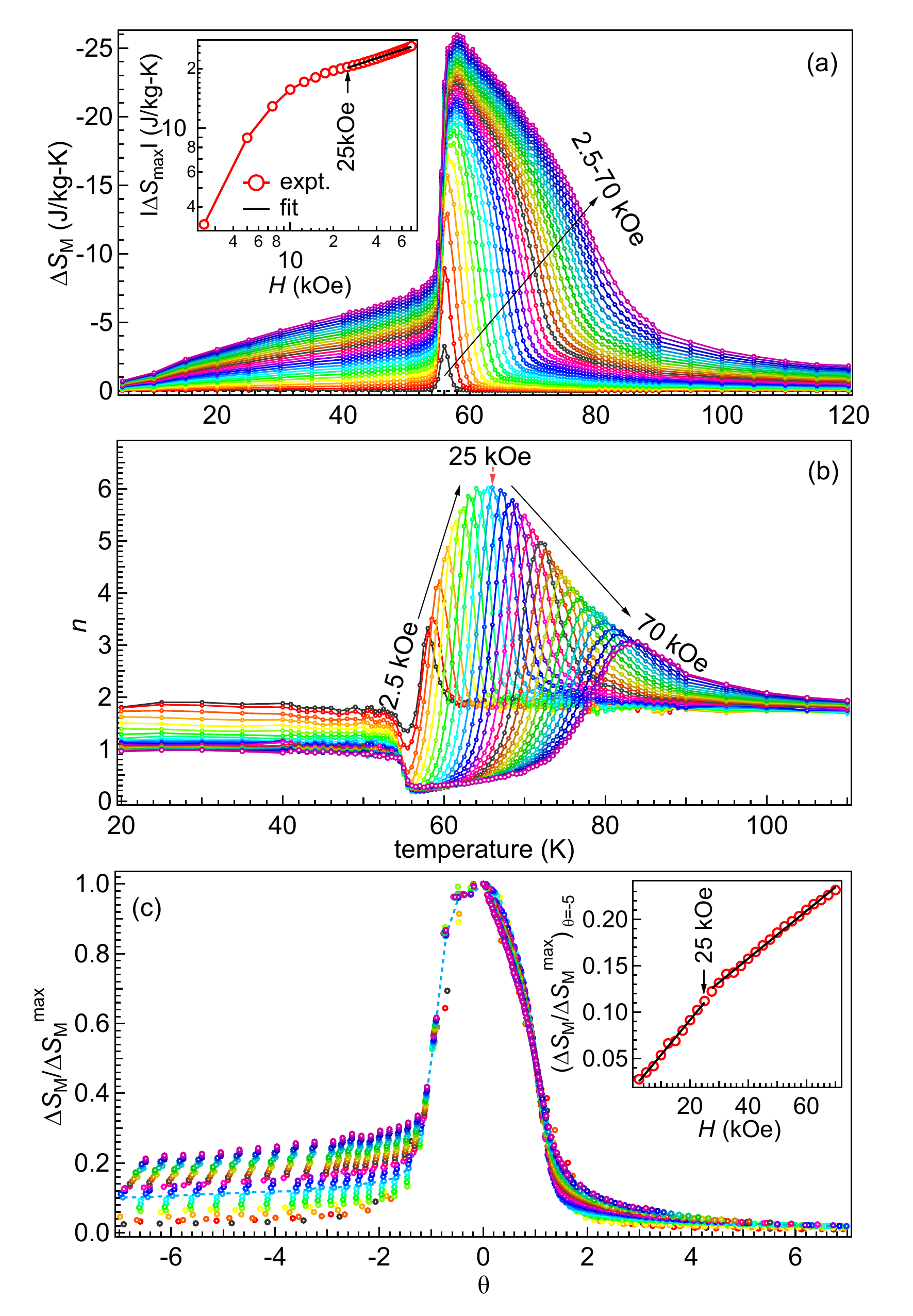}
\caption {(a) Temperature-dependent magnetic entropy change ($\Delta S_{\rm M}$) of Eu$_2$In at different magnetic fields. The inset shows the field dependence of the maximum magnetic entropy change ($\Delta S^{\rm max}_{\rm M}$) on a log-log scale, where the solid black line represents the linear fit for $H > 25$~kOe. (b) Temperature-dependent local field exponent ($n$) of $\Delta S_{\rm M}$ at different fields. (c) Normalized $\Delta S_{\rm M}$ versus scaled temperature curves at different fields, with the dashed line representing the curve for $H = 25$~kOe. The inset shows the field dependence of the vertical dispersion at $\theta = -5$, where the solid black lines represent straight fits to the data below and above $H = 25$~kOe.} 
\label{Fig2}
\end{figure}

\begin{eqnarray} 
 \Delta S_{\rm M}(T,H)=\int_0^H\bigg (\frac{\partial M(T,H)} {\partial T}\bigg)_H dH
\label{delS}
\end{eqnarray}

The $\Delta S_{\rm M}$($T$) curves for Eu$_2$In are shown in Fig. \ref{Fig2}(a) for $\Delta H$ = 2.5--70~kOe, revealing a GMCE with a maximum magnetic entropy change ($\Delta S_{\rm M}^{\rm max}$) of $\approx$~-20~J/kg-K for $\Delta H$ = 20~kOe. The value of $\Delta S_{\rm M}^{\rm max}$ increases rapidly with increasing $\Delta H$ up to $\sim$25~kOe and then rises at a much slower rate at higher fields, as shown in the inset of Fig. \ref{Fig2}(a) on a log-log scale. For a second-order FM–PM transition, $\Delta S_{\rm M}^{\rm max}$ is expected to follow a power-law dependence on the applied magnetic field, $\Delta S_{\rm M}^{\rm max} \propto H^n$ \cite{Franco_APL_06}. In Eu$_2$In, the $\Delta S_{\rm M}^{\rm max}(H)$ curve deviates from this power-law behavior for $H < 25$~kOe, as evidenced by the nonlinearity in the log-log plot presented in the inset of Fig. \ref{Fig2}(a) [see Figs. S1(a, b) of \cite{SI} for more clarity]. In contrast, an almost linear behavior is observed  for $H>$ 25~kOe, suggesting a qualitative change in the magnetic response above this field. A power-law fit to the 25--70~kOe data (solid black line in the inset) yields $n = 0.24$, which is lower than the theoretical value for any established universality class of FM materials \cite{Franco_APL_06}. This suggests that although a notable change in the magnetic behavior of the sample is observed at H$\sim$25~kOe, as predicted in Ref. \cite{Alho_PRB_20}, the overall nature of the phase transition does not transform into a purely second-order type even for $H>$ 25~kOe. 

Further, the $\Delta S_{\rm M}$ ($T$) curves remain nearly symmetric up to $H$ = 25~kOe and then gradually become asymmetric towards the PM region at higher fields [see Fig. \ref{Fig2}(a)]. 
To better understand this, we present local field exponent of entropy, $n$, defined as \cite{Law_NC_18}

\begin{eqnarray} 
n(H, T)= \frac{{\rm d}({\rm ln}|\Delta S_{\rm M}|)} {{\rm d}({\rm ln}H)}
\end{eqnarray}

in Fig. \ref{Fig2}(b) at various magnetic fields. Note that for a second-order FM--PM transition,  $n$ = 1 in the FM state, attains a minimum at $T_{\rm c}$, given by $n$($T_{\rm c}$) = 1 + (1- $\beta^{-1}$)$\delta^{-1}$, where $\beta$ and $\delta$ are the critical exponents, and approaches $n = 2$ in the PM region as expected from Curie-Weiss law \cite{Franco_APL_06}. In contrast, $n > 2$ signifies a first-order transition \cite{Law_NC_18}.  Interestingly, in Eu$_2$In, $n$ initially increases monotonically in the vicinity of $T_{\rm c}$, attains its maximum value of $n\sim$6 at $H$ = 25~kOe and then decreases with further increase in the magnetic field [see Fig. \ref{Fig2}(b)]. This suggests a possible transformation from first-order to second-order character of this FM--PM phase transition in Eu$_2$In at $H$ = 25~kOe. Nevertheless, $n > 2$ persists up to 70~kOe, which indicates that the overall nature of the transition remains first-order up to 70 kOe, but its first-order character likely decreases above 25~kOe. 

To further clarify this, we construct a universal master curve (UMC) of $\Delta S_{\rm M}$($T$) data measured at different magnetic fields by plotting the normalized entropy ($\Delta S_{\rm M}/\Delta S_{\rm M}^{\rm max}$) versus scaled temperature ($\theta$), defined as

\begin{eqnarray} 
\theta= 
\begin{cases} 
 -(T-T_{pk})/(T_{r1}-T_{pk});  T\leqslant T_{\rm c} \\
 (T-T_{pk})/(T_{r2}-T_{pk});   T> T_{\rm c}
\end{cases}
\end{eqnarray} 

where $T_{r1}$ and $T_{r2}$ are the two reference temperatures below and above $T_{\rm c}$, respectively, for a certain value of $h = \Delta S_{\rm M}/\Delta S_{\rm M}^{\rm max}$, and $T_{pk}$ represents the temperature corresponding to $\Delta S_{\rm M}^{\rm max}$ \cite{Franco_APL_06}. The normalized entropy versus scaled temperature for Eu$_2$In is presented in Fig. \ref{Fig2}(c) at different applied magnetic fields for $h = 1/2$. For a second-order FM--PM transition, curves at different fields collapse into a single universal master curve in the entire temperature range, whereas a vertical dispersion is observed for first-order transition below $T_{\rm c}$ \cite{Bonilla_PRB_10}. In the present case, a clear dispersion in the scaled curves is observed as a function of $H$ below $T_{\rm c}$ [see Fig. \ref{Fig2}(c)]. Interestingly, the curves exhibit considerably larger dispersion up to $H \approx 25$~kOe [indicated by the dashed line in Fig. \ref{Fig2}(c)], which is subsequently reduced at higher magnetic fields. To clearly demonstrate this, we plot the field dependence of the normalized entropy at $\theta = -5$ in the inset of Fig. \ref{Fig2}(c). A clear decrease in the dispersion rate is observed around 25~kOe, where the two solid black lines are straight fits to the data below and above $H = 25$~kOe. However, although at a lower rate, the scaled entropy curves still show dispersion up to 70~kOe, which further indicates only a reduction in the first-order-like character of this magnetic transition above 25~kOe, but not a complete transformation to second-order nature, analogous to the behavior of $\Delta S_{\rm M}^{\rm max}$ and $n$, discussed above.

\subsection{\noindent ~Specific heat}

 \begin{figure}
\includegraphics[width=3.5in]{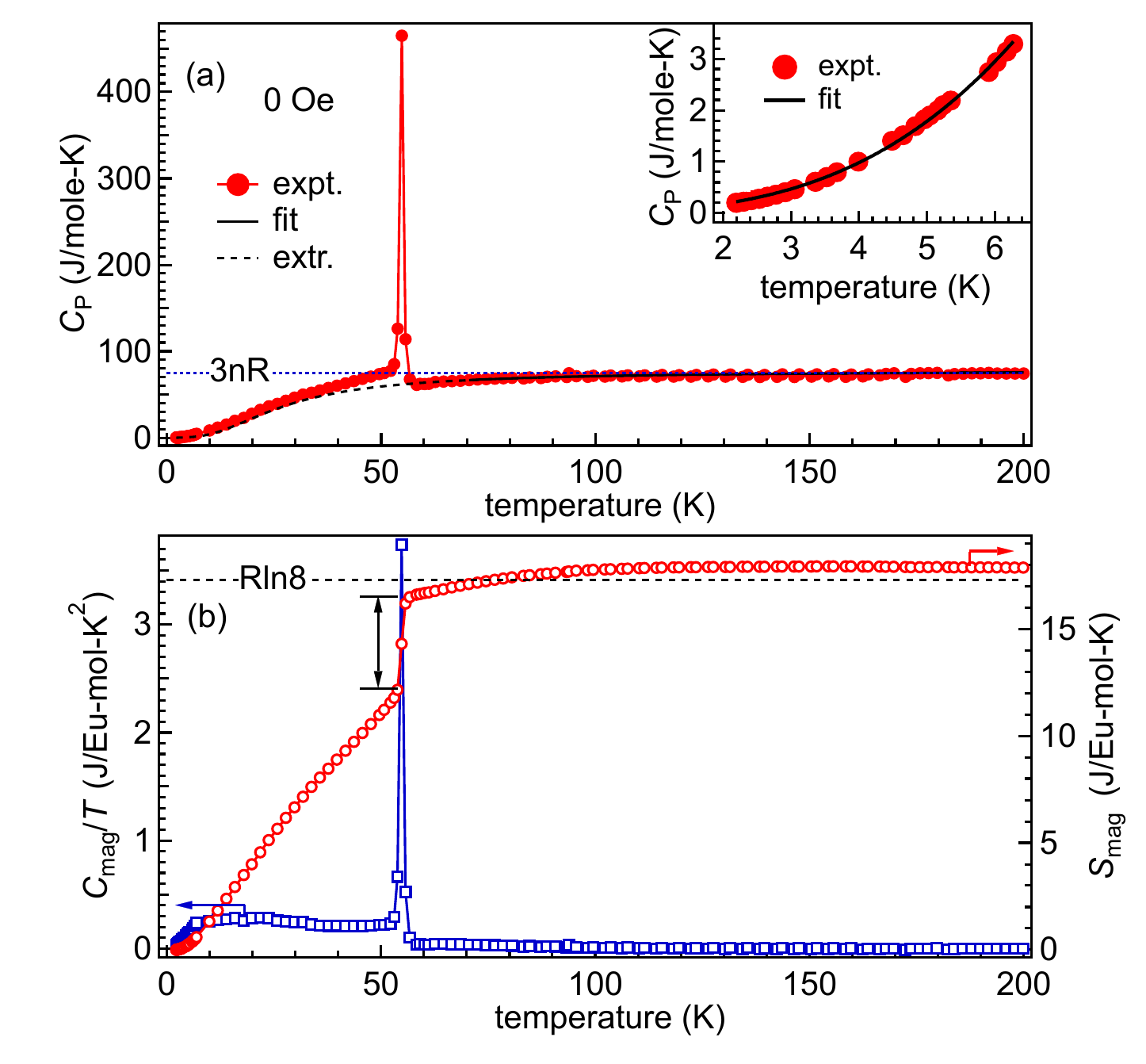} 
\caption {(a) Temperature-dependent zero-field specific heat of Eu$_2$In measured using the standard $2\tau$ relaxation method. The solid and dashed black curves represent the best fit of the data using Eq.~(\ref{Debye}) in the temperature range 70--100~K and the extrapolation of the fitted curve down to 2~K, respectively. The dotted horizontal line represents the classical Dulong--Petit limit of $C_P$ for Eu$_2$In. The inset shows the enlarged view of the low temperature region, where the solid black line represents the best fit using Eq. \ref{HC_LT}. (b) Magnetic contribution to the specific heat, $C_{\rm mag}/T$ (left axis), and the magnetic entropy, $S_{\rm mag}$ (right axis), of Eu$_2$In. The dashed horizontal line represents the theoretical magnetic entropy $S_{\rm mag} = R\ln(2J+1)$ expected for Eu$^{2+}$. The double-sided black arrow indicates the magnetic entropy associated solely with the magnetic transition. }
\label{Fig3}
\end{figure}

The zero-field specific heat ($C_P$) of Eu$_2$In, measured as a function of temperature using the $2\tau$ relaxation method, is shown in Fig.~\ref{Fig3}(a). A sharp, $\delta$-like peak is clearly observed at $T_{\rm c} \approx 55$~K, indicating the first-order nature of the magnetic transition. At low temperatures, the specific heat can be expressed as

\begin{eqnarray}
C_P = \gamma T + \beta T^3 + AT^{3/2}, 
\label{HC_LT}
\end{eqnarray}

where the first, second, and  third terms correspond to electronic, lattice (phonon), and ferromagnetic magnon  contributions, respectively \cite{Anand_PRB_12}. A fit to the $C_P$ data using the above equation in the range $\sim$2--6.5~K, shown by the solid black line in the inset of Fig.~\ref{Fig3}(a), yields  $\gamma = 9.0(7)\ \mathrm{mJ\,mol^{-1}\,K^{-2}}$, $\beta = 11.7(5)\ \mathrm{mJ\,mol^{-1}\,K^{-4}}$, and $A = 24.1(4)\ \mathrm{mJ\,mol^{-1}\,K^{-5/2}}$. Using this value of $\gamma$ and the Debye model for the lattice heat capacity, we fit the specific heat data in the high-temperature paramagnetic regime (70--200~K) by following relation \cite{Kittel_book_05, Kumar_PRB1_25}

\begin{eqnarray}
C_{P} = \gamma T + 9nR\left(\frac{T}{\theta_D}\right)^3
\int_0^{\theta_D/T} \frac{x^4 e^x}{(e^x - 1)^2} dx,
\label{Debye}
\end{eqnarray}

where $n$ is the number of atoms per formula unit (three in the present case), $R$ is the universal gas constant, and $\theta_D$ is the Debye temperature. The best fit of the $C_{P}$ data, shown by the solid black curve in Fig.~\ref{Fig3}(a), gives $\theta_D = 113 \pm 2$~K, which is in good agreement with the value reported for Pr$_2$In \cite{Liu_PRB_24}. The dashed curve in Fig.~\ref{Fig3}(a) represents the extrapolation of the fitted curve down to 2~K, while the dotted horizontal line denotes the classical Dulong--Petit limit, $3nR = 74.8$~J\,mol$^{-1}$\,K$^{-1}$.

 \begin{figure*}
\includegraphics[width=7in]{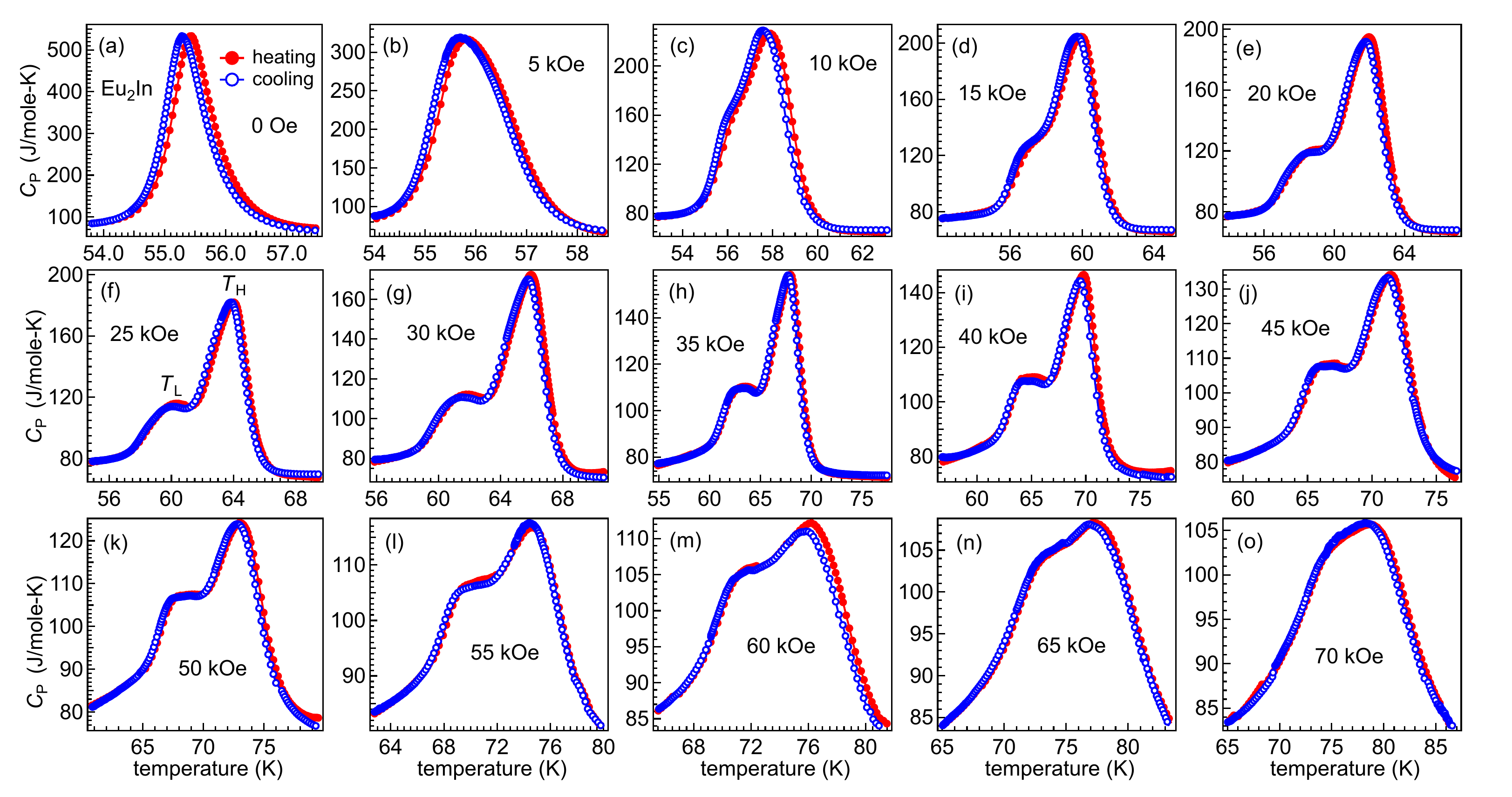} 
\caption {(a--o) The temperature dependent specific heat of Eu$_2$In in the vicinity of magnetic transition at different magnetic fields measured in both heating and cooling modes, using the long heat pulse method. T$_{\rm H}$ and T$_{\rm L}$ in panel (f) represent the temperatures corresponding to the HT and LT peaks, respectively, at 25~kOe.  }
\label{Fig4}
\end{figure*}

The magnetic contribution to the specific heat, $C_{\rm mag}$, is obtained by subtracting the lattice contribution estimated from the nonmagnetic isostructural reference compound Yb$_2$In \cite{Biswas_JPD_26}, by approximating the lattice heat capacity of Eu$_2$In to be equal to that of Yb$_2$In (see Fig.~\ref{Fig11} of the Appendix). The resulting $C_{\rm mag}$ is then used to calculate the magnetic entropy, using 

\begin{eqnarray}
S_{\rm mag}(T) = \int_0^T \frac{C_{\rm mag}(T)}{T} dT.
\label{Smag}
\end{eqnarray}

The $C_{\rm mag}/T$ and $S_{\rm mag}$ are shown on the left and right axes of Fig.~\ref{Fig3}(b), respectively. The dashed horizontal line represents the theoretical magnetic entropy value $S_{\rm mag} = R\ln(2J+1)$ expected for Eu$^{2+}$ with $J = 7/2$. Interestingly, despite the sharp first-order $\delta$-like anomaly in $C_{P}$, only about 25\% of the total Eu$^{2+}$ magnetic entropy is released at the transition itself [as indicated by the double-sided arrow in Fig.~\ref{Fig3}(b)], with the remaining entropy being recovered gradually at lower temperatures. This implies that the magnetic entropy change $\Delta S_{\rm M}$ in Eu$_2$In could, in principle, be enhanced by approximately a factor of four if the full magnetic entropy were utilized, for example by tuning the free-energy landscape through chemical substitution while preserving the large magnetic moment of the ferromagnetic state \cite{Pecharsky_JMMM_2009}.

Note that the standard 2$\tau$ relaxation method for specific heat measurements assumes that (i) the sample’s specific heat remains constant within the small temperature interval raised by a single heat pulse, and (ii) the heating and cooling cycles yield identical results for a given pulse \cite{Hardy_JPCM_09, Lashley_Cryo_43}. However, both the assumptions break down in the case of a sharp ($\delta$-like) first-order transition, leading to inaccuracies in both the position and magnitude of the anomaly \cite{Hardy_JPCM_09, Lashley_Cryo_43}. Therefore, to accurately capture the first-order transition in Eu$_2$In, we apply a long heat pulse just below the transition temperature and analyze heating and cooling cycles separately \cite{Hardy_JPCM_09, Kumar_PRB_24}.  The temperature-dependent specific heat measurements performed using the long heat pulse method under various external magnetic fields are shown in Figs. \ref{Fig4}(a–o).

 \begin{figure}
\includegraphics[width=3.5in]{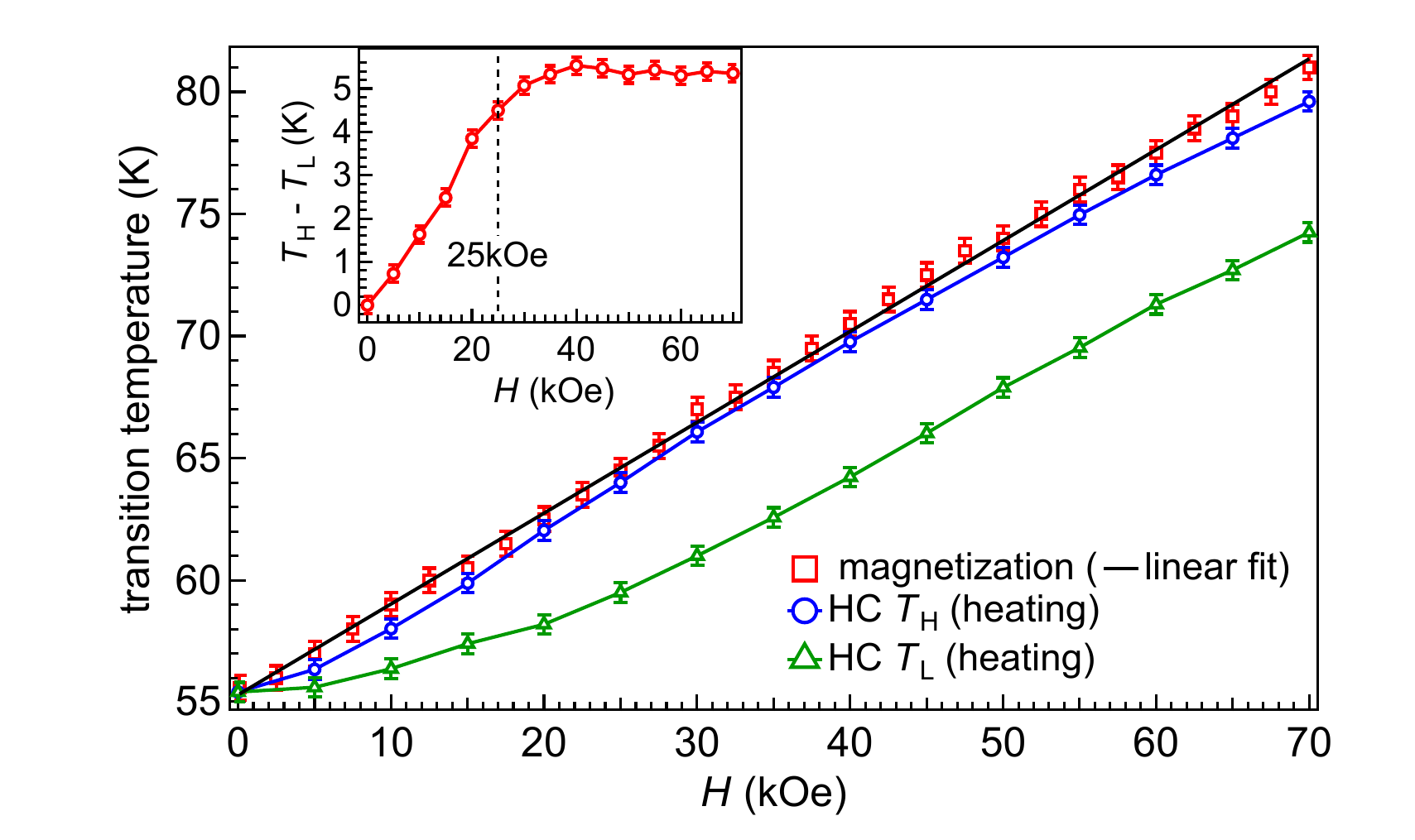} 
\caption {Magnetic-field dependence of the transition temperatures determined from magnetization and specific-heat measurements. The solid black line represents a linear fit to the $T_{\rm c}(H)$ curve, extracted from the magnetization data. The inset shows the field-dependent separation between the two peaks observed in the specific heat curves. }
\label{Fig5}
\end{figure}

 \begin{figure*}
\includegraphics[width=6in]{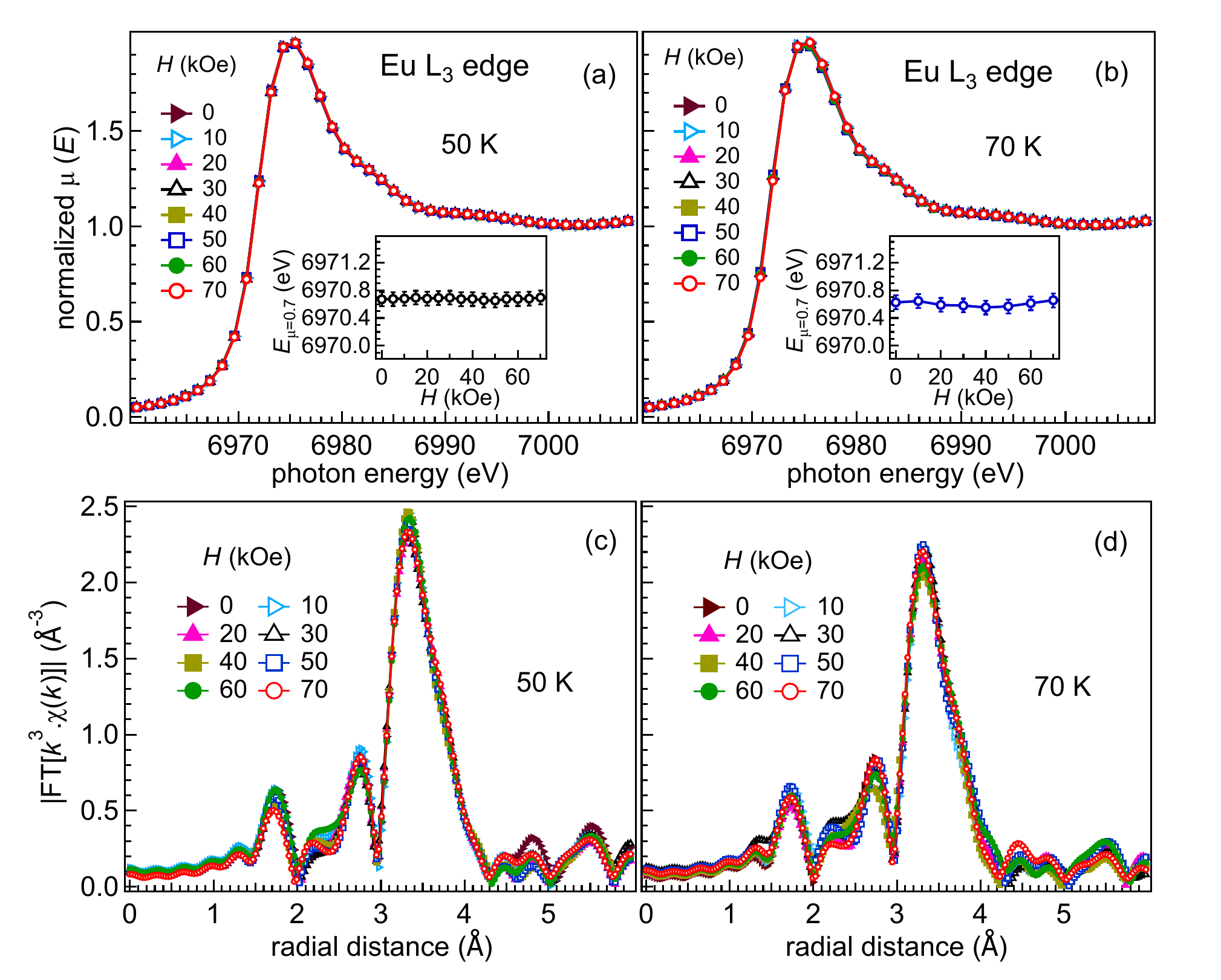} 
\caption {(a, b) Normalized XANES spectra at the Eu $L_{3}$ edge of Eu$_2$In measured under different magnetic fields at 50 and 70~K, respectively. The insets show the energy position of the rising edge at $\mu = 0.7$ as a function of magnetic field. (c, d) Magnitude of the Fourier transform of the $k^{3}\chi(k)$ EXAFS spectra measured at different magnetic fields at 50 and 70~K, respectively. }
\label{Fig6}
\end{figure*}

In the absence of an external magnetic field, a sharp peak is observed in the $C_{P}(T)$ curves with a thermal hysteresis of $\sim0.1$~K [Fig.~\ref{Fig4}(a)]. Interestingly, with increasing magnetic field, this peak evolves into a well-defined doublet, indicating the emergence of an additional transition. A similar field-induced splitting of the sharp zero-field $C_{P}$ peak in Eu$_2$In has also been observed in Ref.~\cite{Guillou_NC_18}. Once the two peaks are sufficiently separated ($H \gtrsim 20$~kOe), the intensity of the low-temperature (LT) peak remains nearly invariant, whereas that of the high-temperature (HT) peak decreases monotonically with further increase in the magnetic field up to 70~kOe [see Fig.~\ref{Fig12} of the Appendix]. However, the magnetic entropy associated with the transitions remains nearly invariant with the external field (see the inset of Fig.~\ref{Fig13} of the Appendix and the discussion therein). To capture the wide temperature range at higher magnetic fields, where the transition becomes significantly broadened ($H \gtrsim 30$~kOe), more than one heat pulse was employed. A reduction in the thermal hysteresis of the $C_{P}(T)$ curves is also observed with increasing magnetic field. It should be noted, however, that the true value of the thermal hysteresis cannot be reliably determined at fields where multiple heat pulses are required to cover the transition region.

Here it is important to emphasize that magnetization measurements reveal only a single transition, manifested as a sharp and symmetric peak in the $dM/dT$ curves [Figs.~\ref{Fig10}(d--f) of the Appendix]. To elucidate the origin of the two anomalies observed in the $C_{P}(T)$ curves, we compare their peak positions with the transition temperature, $T_{\rm c}$ extracted from magnetization measurements [Fig.~\ref{Fig10}(f) of the Appendix], as shown in Fig.~\ref{Fig5}. The specific-heat peak positions were determined by fitting the data using a sum of two Voigt functions [Figs.~S2(a--c) of Ref.~\cite{SI}]. Notably, the position of the HT peak ($T_{\rm H}$) closely follows $T_{\rm c}$ obtained from magnetization, indicating its magnetic origin, which is further supported by the systematic reduction of its magnitude with increasing field (Fig. \ref{Fig12} of the Appendix).

 \begin{figure*}
\includegraphics[width=7in]{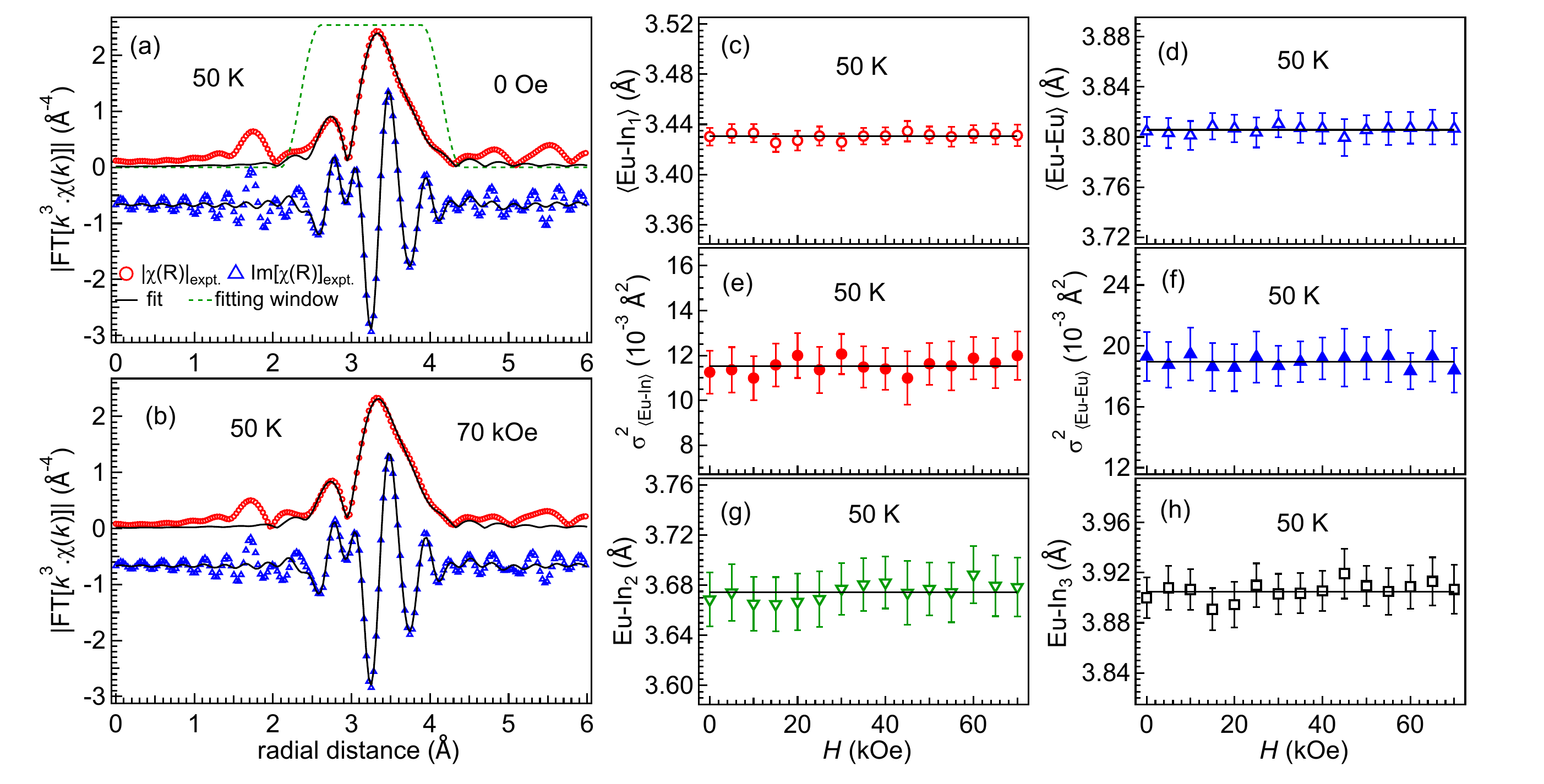} 
\caption {(a, b) Fits to the $|\chi(R)|$ and corresponding $\mathrm{Im}[\chi(R)]$ spectra within the window indicated by the dotted green line [in panel (a)] at magnetic fields of 0 and 70~kOe, respectively, at 50~K. Each $\mathrm{Im}[\chi(R)]$ spectrum is shifted downward for clarity. (c, d) Magnetic-field dependence of the average Eu–In$_1$ and Eu–Eu scattering path distances, respectively. (e, f) Corresponding Debye-Waller factors. (g, h) Magnetic-field dependence of the Eu–In$_2$ and Eu–In$_3$ scattering path distances. Solid black lines indicate the field independent behavior of the various scattering distances and their corresponding Debye-Waller factors.}
\label{Fig7}
\end{figure*}

In contrast, no corresponding feature is observed in the magnetization data near the LT peak of the $C_{P}(T)$ curves, indicating that this anomaly is not directly associated with a magnetic phase transition. A natural interpretation is that the LT anomaly originates from a lattice-related instability that becomes progressively decoupled from the magnetic transition under an applied magnetic field, consistent with the almost field independence of its magnitude [Fig.~\ref{Fig12}]. However, the systematic shift of both the LT and HT peaks to higher temperatures with increasing magnetic field argues against a purely structural transition that is entirely independent of magnetism. Instead, this behavior points to the presence of finite magnetoelastic coupling, whereby the applied magnetic field indirectly influences the lattice degrees of freedom. In this context, a field-induced structural response, such as a magnetostriction-driven effect, remains a plausible scenario.

A key observation is that the separation between the two peaks in the specific-heat curves increases rapidly with magnetic field up to $\sim$25~kOe and then remains nearly constant at higher fields, as shown in the inset of Fig.~\ref{Fig5}, while both transitions broaden with increasing field. This decoupling of the two transitions near $H \sim 25$~kOe may be responsible for the apparent suppression of the first-order--like character of the transition above this critical field, as inferred from the magnetization measurements discussed above. A comparison of the $C_{P}$ data obtained using the standard 2$\tau$ relaxation method and the long heat-pulse method at 35~kOe  shows excellent reproducibility (Fig.~S3 of Ref.~\cite{SI}), confirming that the sharp zero-field anomaly indeed splits into two distinct features under applied magnetic field.

\subsection{\noindent ~X-ray absorption spectroscopy}

Magnetic field dependent x-ray absorption spectroscopy (XAS) measurements were performed to probe possible field-induced changes in the local crystal structural and/or electronic environment of Eu$_2$In. Figures~\ref{Fig6}(a) and \ref{Fig6}(b) show the normalized XANES spectra measured at the Eu $L_{3}$ edge under different magnetic fields at 50~K ($< T_{\rm c}$) and 70~K ($> T_{\rm c}$), respectively. An intense feature around 6976~eV is attributed to the $2p_{3/2} \rightarrow 5d$ transition of Eu$^{2+}$ \cite{Jiang_PRB_17, Neto_PRL_12, Tan_JACS_16}, indicating a predominantly divalent state of Eu in the sample, consistent with the magnetization measurements discussed above and with previous reports \cite{Guillou_NC_18, Ryan_AIP_19}. More importantly, the valence state of Eu remains invariant with applied magnetic field at both temperatures. Indeed, the entire XANES line shape remains unchanged with $H$, including the position of the rising-edge region, as illustrated in the insets of Figs.~\ref{Fig6}(a, b) at $\mu = 0.7$, where $\mu$ denotes the normalized absorption coefficient. These results demonstrate that the electronic structure of Eu$_2$In is robust against the application of an external magnetic field.

To further investigate the local coordination environment around Eu atoms, we performed EXAFS measurements at the Eu $L_{3}$ edge [Figs.~S4(a, b) of Ref.~\cite{SI}]. Figures~\ref{Fig6}(c, d) show the magnitude of the Fourier transform of the $k^{3}\chi(k)$ spectra [Figs.~S4(c, d) of Ref.~\cite{SI}] measured at different magnetic fields for $T=50$ and 70~K, respectively. The $\chi(R)$ spectra were obtained using a Hanning window over the $k$ range 2.5--11.5~\AA$^{-1}$ with a background cutoff of  $R_{\mathrm{bkg}}=1.8$~\AA. The dominant feature at $R \sim 3.5$~\AA\ (phase uncorrected) arises primarily from nearest-neighbor Eu--In single-scattering paths. Upon closer inspection, this peak exhibits a noticeable asymmetry toward higher radial distance, indicating additional contributions from Eu--Eu correlations and/or longer Eu--In scattering paths. In the orthorhombic $Pnma$ structure, Eu atoms occupy two crystallographically distinct sites (Eu1 and Eu2). Each Eu atom is coordinated by five In atoms (within 4.5~\AA), resulting in four distinct Eu--In bond distances. Because the two shortest Eu--In distances (one singly coordinated and one twofold coordinated) are close and cannot be reliably resolved within the present EXAFS resolution, we modeled them using an average distance $\langle \mathrm{Eu\!-\!In}_{1}\rangle$, together with independent $\langle \mathrm{Eu\!-\!In}_{2}\rangle$ and $\langle \mathrm{Eu\!-\!In}_{3}\rangle$ distances, where $\mathrm{In}_{n}$ denotes the $n^{\mathrm{th}}$ nearest-neighbor In atom. To avoid overparameterization, a common Debye--Waller (DW) factor was used for all Eu--In paths.

 \begin{figure*}
\includegraphics[width=7in]{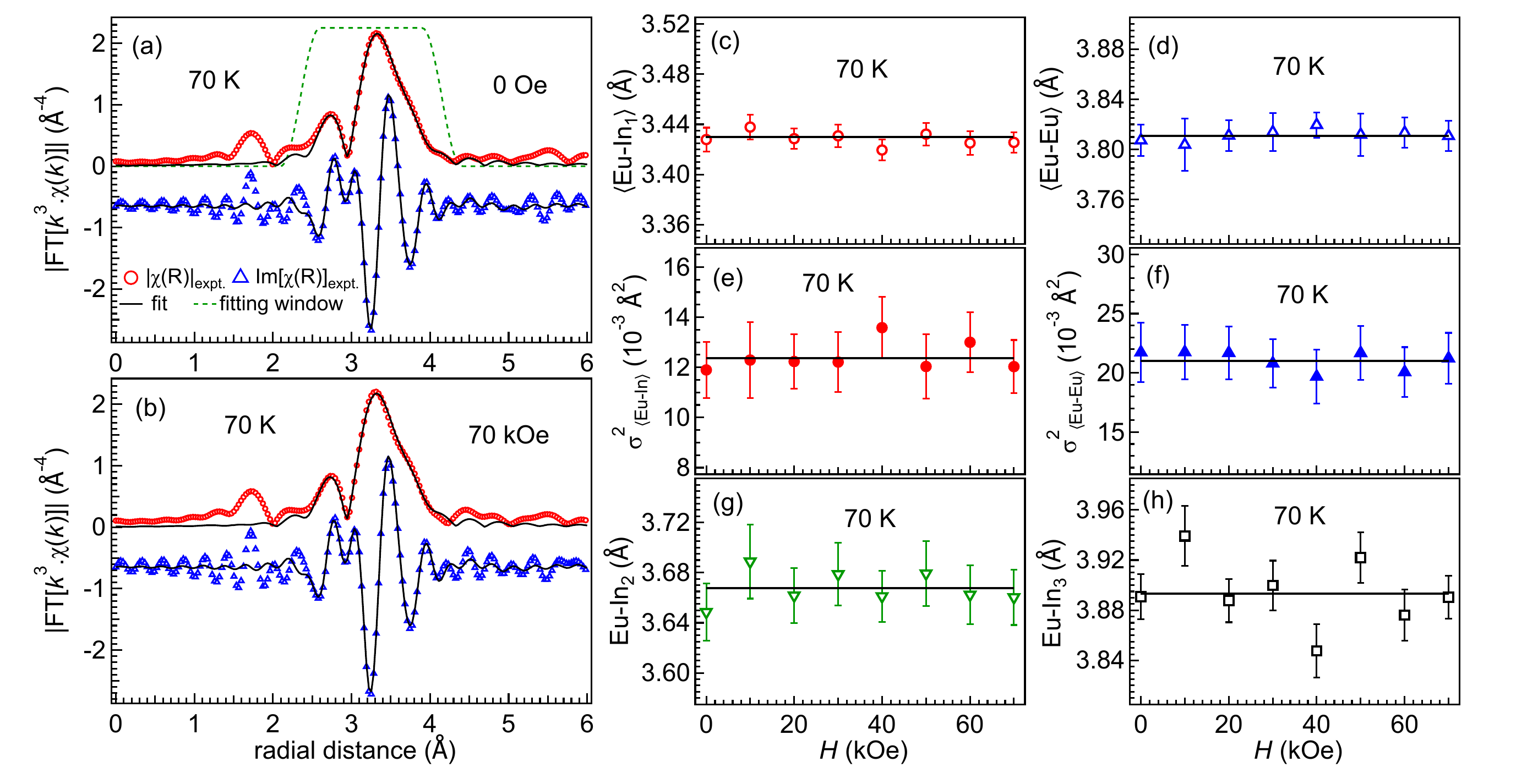} 
\caption {(a, b) Fits to the $|\chi(R)|$ and corresponding $\mathrm{Im}[\chi(R)]$ spectra within the window indicated by the dotted green line [in panel (a)] at magnetic fields of 0 and 70~kOe, respectively, at 70~K. Each $\mathrm{Im}[\chi(R)]$ spectrum is shifted downward for clarity. (c, d) Magnetic-field dependence of the average Eu–In$_1$ and Eu–Eu scattering path distances, respectively. (e, f) Corresponding Debye-Waller factors. (g, h) Magnetic-field dependence of the Eu–In$_2$ and Eu–In$_3$ scattering path distances. Solid black lines indicate the field independent behavior of the various scattering distances and their corresponding Debye-Waller factors.}
\label{Fig8}
\end{figure*}

Further, both Eu1 and Eu2 sites are coordinated by eight Eu atoms, comprising five distinct Eu--Eu distances. Since the spread among these Eu--Eu distances is below the resolution of the present EXAFS data ($\lesssim 0.17~\text{\AA}$), we modeled them using a single average Eu--Eu bond length $\langle \mathrm{Eu\!-\!Eu}\rangle$, with equal weighting of contributions from both Eu sites. The $\chi(R)$ spectra were first fitted at $T=10$~K in zero field (not shown), and the amplitude reduction factor,  $S_{0}^{2}$ and energy shift, $\Delta E_{0}$ were subsequently fixed for all other temperatures and magnetic fields. Fits were performed over the range $R=2.35$--4.15~\AA\ [green dashed curves in Figs.~\ref{Fig7}(a) and \ref{Fig8}(a)]. Representative fits at 0 and 70~kOe are shown in Figs.~\ref{Fig7}(a, b) for 50~K and in Figs.~\ref{Fig8}(a, b) for 70~K (see Fig.~S5 in \cite{SI} for the contributions from individual  scattering paths). The magnetic-field dependence of the extracted parameters is summarized in Figs.~\ref{Fig7}(c--h) and \ref{Fig8}(c--h). In all fits, $S_{0}^{2}$ was fixed at 0.793, and the resulting $R$-factor (goodness of fit) lies in the range 0.004--0.006 at both temperatures.

Interestingly, the extracted Eu--In and Eu--Eu bond lengths, as well as their corresponding DW factors, exhibit no statistically significant field dependence up to 70~kOe at both 50~K (below $T_{\rm c}$) and 70~K (above $T_{\rm c}$). Since the DW factor captures contributions from both thermal and static disorder, the absence of any field-induced variation indicates that the applied magnetic field does not introduce additional local disorder, nor does it drive a measurable local distortion (e.g., magnetostriction) in the Eu coordination environment. While extremely subtle structural changes below the sensitivity of the present EXAFS measurements ($\approx$ 0.01--0.02~\AA) cannot be completely excluded, our results suggest that the second anomaly observed in the specific heat is unlikely to be associated with a magnetic-field-driven local structural modification in Eu$_2$In.

\section{\noindent ~Discussion}

The change in the behavior of magnetocaloric properties of Eu$_2$In, namely, deviation of the $\Delta S_{\rm M}^{\rm max}(H)$ curve, systematic variation of the local power exponent $n$, and evolution of the UMC around 25~kOe, points toward a modification in the nature of the transition at this critical field. At the same time, the occurrence of an unphysically small value of the power exponent $n$, the overshoot of the local exponent above $n = 2$, and the persistence of dispersion in the UMC even for $H = 25$–70~kOe indicate that the overall transition retains its first-order character. It is important to note that $\Delta S_{\rm M}$ is a cumulative quantity, and its values at higher fields necessarily incorporate contributions from the low-field region [see Eq.~(\ref{delS})]. As a result, even if the transition evolves toward second-order behavior for $H > 25$~kOe, the imprint of the low-field first-order character would persist in the $\Delta S_{\rm M}$ curves at higher fields, giving rise to first-order-like features in all derived quantities.

\begin{figure}
\includegraphics[width=3.5in]{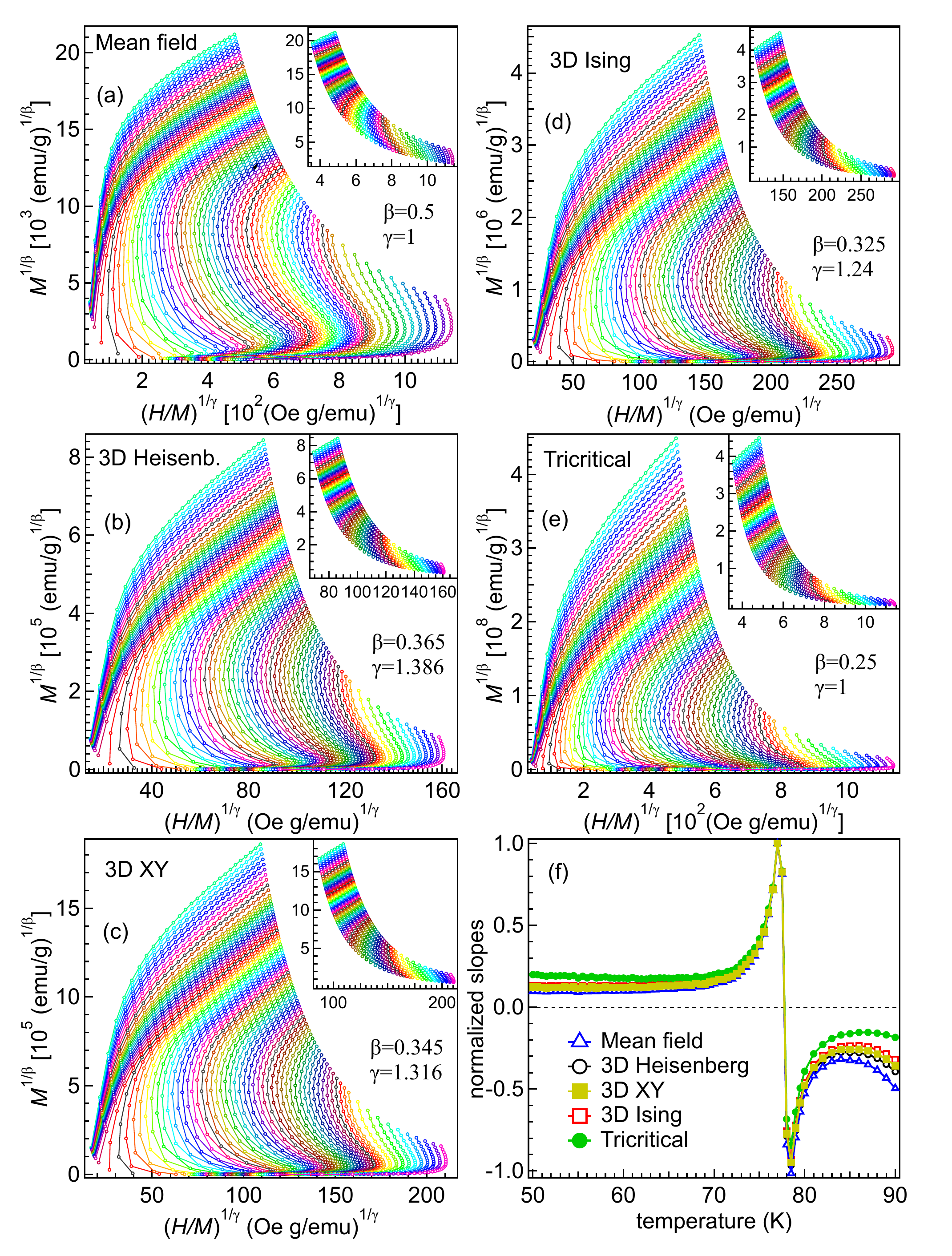} 
\caption {(a-e) Modified Arrott plots (MAPs) [$M^{1/\beta}$ vs $(H/M)^{1/\gamma}$] of Eu$_2$In in the temperature range 40-90 K for different universality classes. Insets show an enlarged view of the high-field (50-70 kOe) region. (f) Normalized slopes of the MAPs obtained from 60-70 kOe field range for different universality classes. }
\label{Fig9}
\end{figure}

To test this scenario, we apply the Banerjee criterion, which states that negative and positive slopes of Arrott plots ($H/M$ versus $M^{2}$) in the high-field region correspond to first-order and second-order transitions, respectively \cite{Arrott_PRL_67, Banerjee_PL_64}. Figures~\ref{Fig9}(a–e) display the modified Arrott plots (MAPs) [$M^{1/\beta}$ versus $(H/M)^{1/\gamma}$] for Eu$_2$In, constructed for different universality classes using their corresponding critical exponents ($\beta$ and $\gamma$) \cite{Guillou_PRB_80}. If the transition indeed crossed over to second-order at high fields, the MAPs would approach straight lines in the high-field regime for the correct choice of $(\beta, \gamma)$ \cite{Sarkar_PRB_08, Roble_PRB_04, Kumar_PRB_25}. In contrast, none of the tested universality classes yield linear behavior in the high-field region. The insets of Figs. \ref{Fig9}(a--e), highlighting the 50--70~kOe  field range, reveal a pronounced curvature that persists up to the highest fields investigated, confirming the robustness of the first-order character. Moreover, the slope of the MAPs in this regime changes from positive to negative with increasing temperature. To quantify this behavior, linear fits were performed in the 60-70~kOe range [see Figs.~S6(a–e) of Ref.~\cite{SI}], and the normalized slope is plotted in Fig.~\ref{Fig9}(f). A distinct crossover from positive to negative slope occurs around 78~K for all universality classes, consistent with first-order-like behavior in which the critical exponents diverge at $T_{\rm c}$. These results unambiguously demonstrate that the first-order nature of the transition in Eu$_2$In remains intact at least up to 70~kOe.

The splitting of the specific-heat peak under an external magnetic field, together with its correlation with the transition temperature observed in magnetization, and the field-dependent variation in the magnitude of the two peaks, suggest that a field-induced subtle change in the crystal structure of Eu$_2$In could be responsible for the LT peak in the $C_{P}$ curves. However, at the same time, magnetic-field-dependent EXAFS measurements rule out any detectable field-induced local structural changes in the sample within the experimental resolution. Recently, Hardy $et$ $al.$ proposed a mathematical model describing a “two-step” process in the first-order transformation of giant magnetocaloric materials \cite{Hardy_AM_22}. In this framework, the transformation initiates and terminates through abrupt changes in the transformed fraction, accompanied by large exchanges of latent heat  \cite{Hardy_AM_22}. In the present case, the external field may induce a similar two-step-like process in Eu$_2$In. Here it is important to mention that the field-induced splitting in the $C_{P}(T)$ curves of Eu$_2$In is fully reversible, where the doublet reverts back to a singlet upon removal of the external field.  We note that different sample geometries were used in the present measurements, namely an irregular bulk piece for magnetization, a thin rectangular plate for specific-heat measurements, and a powdered sample for XAS. Consequently, demagnetization effects, which lead to differences between the applied and internal magnetic fields, may vary among the different measurement techniques, particularly in the low-field regime.  However, this effect is not expected to significantly influence the overall field dependence or the main conclusions of this study. A related two-stage transition has also been reported in Nd$_2$In, where the longitudinal strain first increases with magnetic field and then decreases at higher fields during its first-order transition at 108 K \cite{Liu_APL_21}. The authors correlate this two-stage transition in Nd$_2$In with theoretical calculations for Eu$_2$In, where the free energy with respect to FM order exhibits two minima \cite{Liu_APL_21, Tapia_PRB_20}. At this stage, we believe that more comprehensive investigations such as magnetic field dependent diffraction \cite{Pecharsky_PRL_03, Mudryk_PRL_10} and/or a simultaneous probes of magnetic and structural responses \cite{Aubert_IEEE_22, Karpenkov_APR_23, Karpenkov_PRA_20} are required to clarify the origin of the two closely spaced peaks observed in the $C_{P}(T, H)$ curves of Eu$_2$In.

\begin{figure*}
\includegraphics[width=1\textwidth]{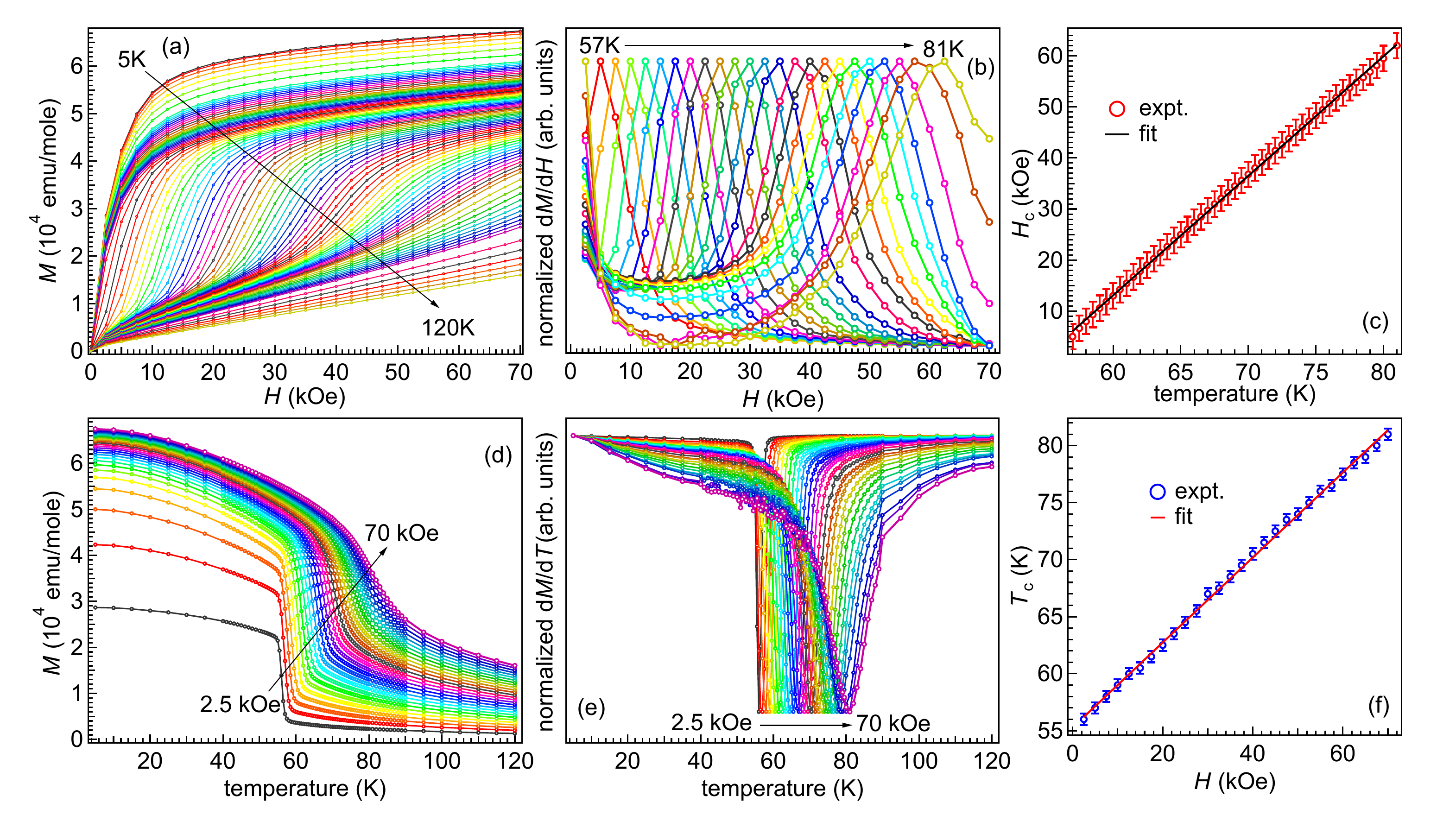} 
\caption {(a) Virgin magnetization isotherms of Eu$_2$In measured from 0 to 70~kOe at different temperatures across $T_{\rm c}$. (b) Normalized field derivative of the magnetization in the temperature range 57--81~K. (c) Temperature dependence of the critical field $H_{\rm c}$; the black solid line represents a linear fit to the data. (d) Temperature-dependent magnetization at different applied fields, extracted from the virgin magnetization isotherms. (e) Normalized temperature derivative of the magnetization at different fields. (f) Field dependence of the transition temperature; the red solid line represents a linear fit to the data.}
\label{Fig10}
\end{figure*}

\section{\noindent ~Conclusion}

In summary, we have investigated the temperature- and magnetic-field-dependent magnetization, specific heat, and local crystal structure of Eu$_2$In across its first-order ferromagnetic--paramagnetic transition. The field dependence of $\Delta S_{\rm M}^{\rm max}$ follows a power-law behavior for $H \gtrsim 25$~kOe, but deviates at lower fields. The local field exponent $n$ increases near $T_{\rm c}$ up to $\sim$25~kOe and then decreases with further increase in field, which, together with the reduced dispersion in the universal master curve of $\Delta S_{\rm M}(H,T)$ around 25~kOe, suggests a tendency of the transition to evolve toward second-order character above this critical field. Nevertheless, quantitative analyses of the magnetocaloric properties and modified Arrott plots unambiguously confirm that the overall first-order nature of the transition persists at least up to 70~kOe. While this behavior contrasts with theoretical predictions of a field-induced first- to second-order transition above $\sim$25~kOe \cite{Alho_PRB_20}, it certainly indicates a subtle modification of the transition mechanism at higher fields. Specific-heat measurements reveal a clear field-induced splitting of the sharp, $\delta$-like anomaly into a doublet. A comparison between the $C_P$ peak positions and the $T_{\rm c}$ obtained from magnetization, together with the field-dependent changes in their relative amplitudes, suggests a nonmagnetic origin for the low-temperature peak. However, at the same time, EXAFS measurements show no detectable changes in the local crystal structure, ruling out a field-induced structural distortion in the sample within the experimental resolution. Taken together, these results point to a field-driven transformation of the transition in Eu$_2$In from a one-step to a two-step process.

\subsection{Acknowledgments}

This work was performed at Ames National Laboratory and was supported by the Division of Materials Science and Engineering of the Office of Basic Energy Sciences, Office of Science of the U.S. Department of Energy (DOE). Ames National Laboratory is operated for the U.S. DOE by Iowa State University of Science and Technology under Contract No. DE-AC02-07CH11358. The XAS measurements were conducted at the Center for High-Energy X-ray Sciences (CHEXS), which is supported by the National Science Foundation (BIO, ENG and MPS Directorates) under award DMR-2342336. T. A. Tyson is supported by NSF Award DMR-2313456. Work at Argonne National Laboratory was supported by the U.S. DOE Office of Science-Office of Basic Energy Sciences, under Contract No.DE-AC0206CH11357.

\section*{APPENDIX}

\setcounter{subsection}{0}
\renewcommand{\thesubsection}{\arabic{subsection}}

\subsection{Additional magnetization data}

Figure \ref{Fig10}(a) shows the virgin magnetization isotherms of Eu$_2$In measured from 0 to 70~kOe at different temperatures across $T_{\rm c}$.  For $T \leqslant T_{\rm c}$, the curves show the saturating behavior similar to that observed at 2~K [Fig. \ref{Fig1}(d)]; however for $T > T_{\rm c}$, the magnetic moment increases almost linearly for the lower applied magnetic fields and then show as step-like jump with further increase in the magnetic field [see Fig. \ref{Fig10}(a)]. The critical magnetic field ($H_{\rm c}$) for this jump in the magnetization, i.e., the magnetic field required to ferromagnetically align the spins above $T_{\rm c}$, increases with the temperature. In order to clearly present this, we show the normalized field derivative of the magnetization in Fig. \ref{Fig10}(b) for $T > T_{\rm c}$, where a monotonic shift in the peak position can be clearly observed to the higher field with increasing sample temperature.  In Fig.  \ref{Fig10}(c), we plot this peak position as a function of temperature, which shows that $H_{\rm c}$ increases linearly with temperature for $T > T_{\rm c}$. Here, we perform the Gaussian fitting of the peaks in the vicinity of the transition to precisely estimate the peak position. The straight fit of the experimental data, as shown by the black solid line in Fig.  \ref{Fig10}(c), gives a significant shift of 2.35~kOe/K. \par

The temperature dependent magnetization extracted from these virgin magnetization isotherms is presented in Fig. \ref{Fig10}(d) at the different magnetic fields. With increase in the magnetic field, the transition becomes broader with a monotonic shift in the transition temperature to the higher values. The normalized d$M$/d$T$ is plotted as a function of temperature in  Fig. \ref{Fig10}(e) at different fields to show the field evolution of the $T_{\rm c}$. The field dependence of $T_{\rm c}$ is shown in Fig. \ref{Fig10}(f), which shows the linear behavior up to 70~kOe and a straight fit of the data gives a shift of 0.37~K/kOe in the transition temperature. This linear dependence of the transition temperature of Eu$_2$In on the external field makes it a suitable candidate for the device application in a wide range of the working temperature.

\subsection{Additional specific heat data}

\begin{figure*}
\centering
\includegraphics[width=1\textwidth]{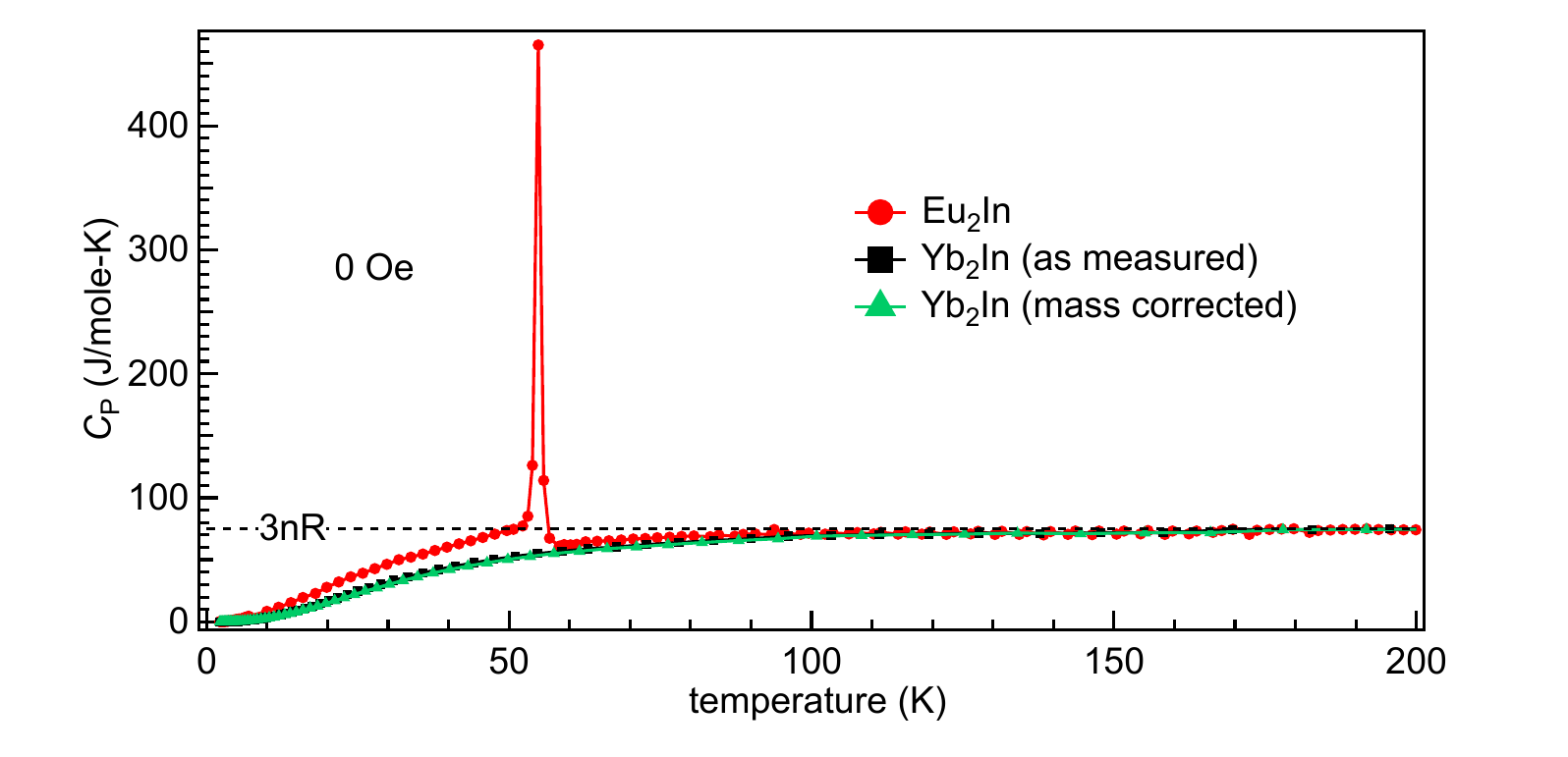}
\caption {Temperature-dependent specific heat of Eu$_2$In and its isostructural nonmagnetic reference compound Yb$_2$In measured in zero magnetic field using the $2\tau$ relaxation method. The Yb$_2$In data are shown both as measured and after mass correction to account for the difference in atomic masses. The horizontal dashed line represents the classical Dulong--Petit limit.}
\label{Fig11}
\end{figure*}

Figure~\ref{Fig11} shows the zero-field temperature dependent specific heat data of Eu$_2$In (solid red circles). To account for the lattice (phonon) contribution to the specific heat of Eu$_2$In, we measured the $C_P(T)$ data of its isostructural nonmagnetic reference compound Yb$_2$In \cite{Biswas_JPD_26}. Since Eu$_2$In and Yb$_2$In have different formula unit masses, a correction for the mass difference was considered. Within the Debye approximation, the phonon heat capacity depends on the reduced temperature $T/\theta_D$ [see Eq.~\ref{Debye}]. The Debye temperature $\theta_D$ depends on the formula unit mass $M$ as $\theta_D \propto M^{-1/2}$ \cite{Anand_PRB_15}. Consequently, the measured temperature scale of Yb$_2$In was rescaled according to the relation \cite{Anand_PRB_15}

\begin{equation}
T^* = T \left( \frac{M_{\mathrm{Yb_2In}}}{M_{\mathrm{Eu_2In}}} \right)^{1/2}.
\end{equation}

The as-measured and mass-corrected $C_P(T)$ data of Yb$_2$In are shown in Fig.~\ref{Fig11}. No significant difference between the two datasets is observed in the present case. Therefore, the as-measured $C_P(T)$ data of Yb$_2$In are used to account for the lattice heat capacity of Eu$_2$In in all subsequent analyses \cite{Anand_PRB_09, Anand_PRB_12}.

\begin{figure}
\centering
\includegraphics[width=0.5\textwidth]{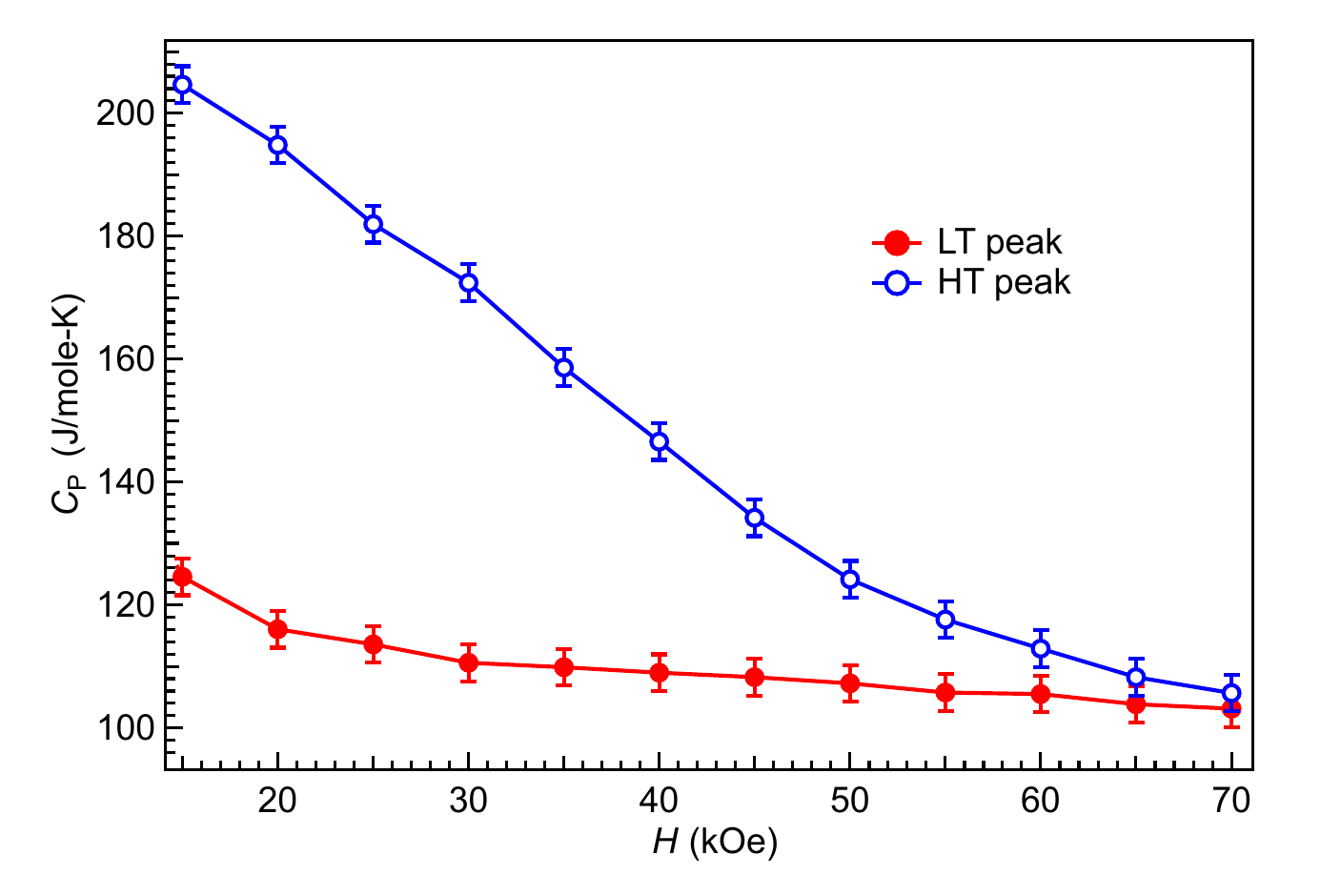}
\caption {Magnetic field dependence of the peak values of the low- and high-temperature anomalies in the specific heat of Eu$_2$In.}
\label{Fig12}
\end{figure}

\begin{figure*} 
\includegraphics[width= 1 \textwidth]{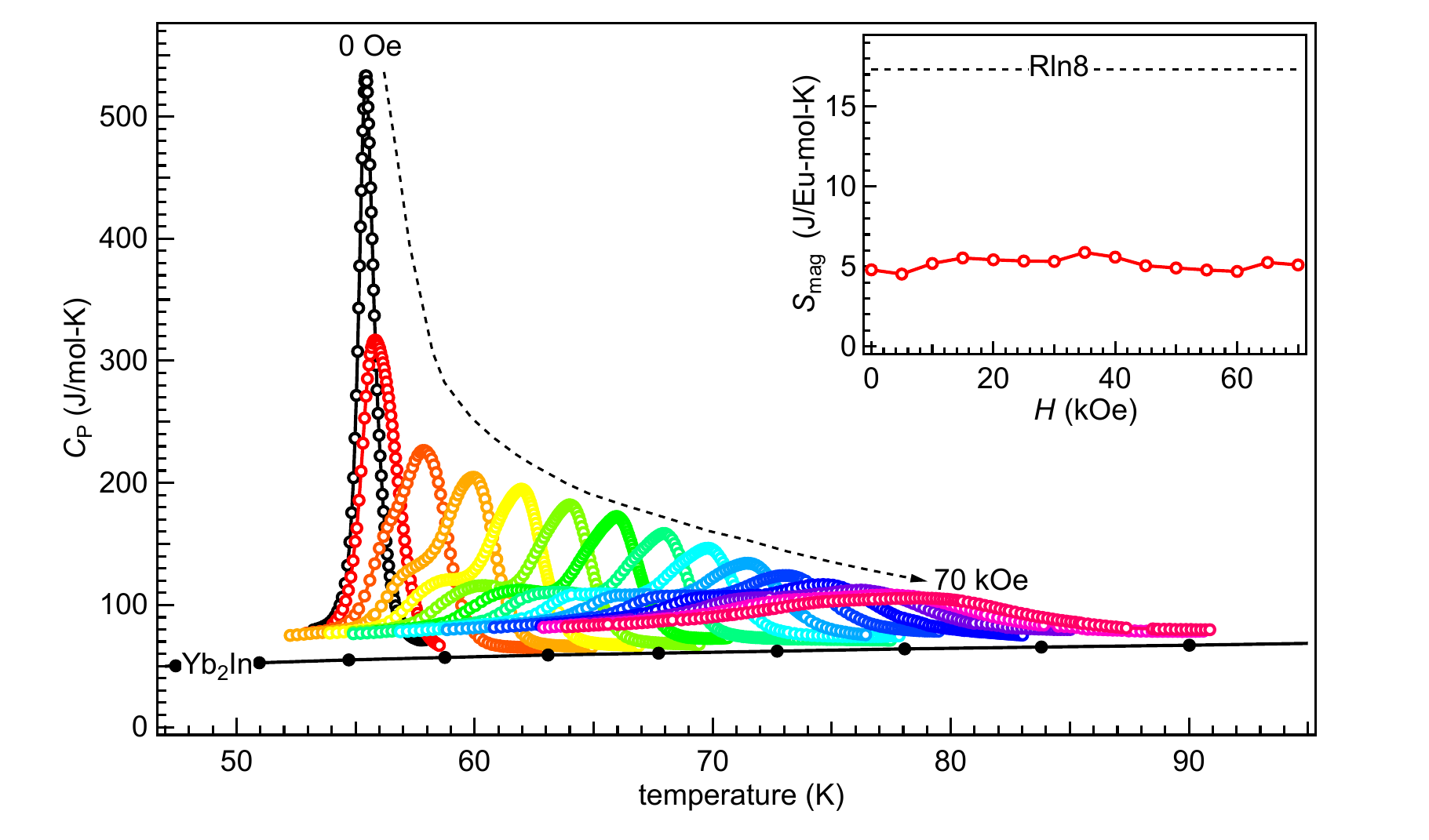} 
\caption {Temperature dependent specific heat $C_P(T)$ of Eu$_2$In in the vicinity of the magnetic transition measured using the long heat-pulse method at different applied magnetic fields. The black solid circles represent the $C_P(T)$ data of the nonmagnetic reference compound Yb$_2$In. Inset shows the magnetic entropy $S_{\mathrm{mag}}$ associated with the transition as a function of magnetic field. The horizontal dashed line denotes the theoretical entropy $R\ln(2J+1)$ expected for Eu$^{2+}$ ($J=7/2$).}
\label{Fig13}
\end{figure*}

Figure~\ref{Fig12} shows the magnetic-field dependence of the peak magnitudes of the two anomalies observed in the specific-heat data of Eu$_2$In (see Fig.~\ref{Fig4}), after deconvolution for $H \geqslant 15$~kOe. The low-temperature (LT) peak is almost field invariant, whereas the high-temperature (HT) peak decreases monotonically with increasing field and approaches the LT value near $H \approx 70$~kOe. The field invariance of the LT anomaly suggests a nonmagnetic (likely structural) origin of this feature.

Figure~\ref{Fig13} shows the temperature dependence of the specific heat of Eu$_2$In near the magnetic transition under different applied magnetic fields, measured using the long heat-pulse method in heating mode. The $C_P(T)$ data of the isostructural nonmagnetic reference compound Yb$_2$In, shown by the black solid curve, are used to estimate the lattice contribution to the specific heat of Eu$_2$In. The magnetic entropy calculated by considering this lattice contribution is shown in the inset of Fig.~\ref{Fig13}. It is observed that the magnetic entropy associated with the transition is nearly field independent and is significantly lower than the theoretical value expected for Eu$^{2+}$, as indicated by the horizontal dashed line in the inset of Fig.~\ref{Fig13}. Notably, the zero-field magnetic entropy associated with the transition is close to that reported in Ref.~\cite{Guillou_NC_18} ($\sim$5.7~J/Eu-mol-K), where a step-function approximation was used to estimate the lattice background in the transition region, despite the significantly larger peak value of $C_P$ observed in Ref.~\cite{Guillou_NC_18}. This suggests that the transition in the present case is broader, possibly reflecting differences in compositional homogeneity between the two samples \cite{Jones_JAP_12}.

\end{document}